\documentclass[oneside, a4paper, onecolumn, 11pt]{article}
\usepackage[left=2.5cm,top=2.5cm,bottom=2.5cm,right=2.5cm]{geometry}
\usepackage{amsmath}
\usepackage{graphicx}
\usepackage{fancyhdr}
\usepackage{calc}
\usepackage{amssymb}
\usepackage[compress,numbers,square]{natbib}
\usepackage{setspace}
\usepackage{amsfonts}
\usepackage{commath}
\usepackage[innercaption]{sidecap}
\usepackage[framemethod=TikZ]{mdframed}
\usepackage{hyperref}
\usepackage{parskip}
\usepackage{pifont}
\usepackage{changepage}
\usepackage{sectsty}
\usepackage[labelfont={bf},font={footnotesize,sf}]{caption}
\sectionfont{\color{blue}}  
\subsectionfont{\color{blue}}  
\subsubsectionfont{\color{blue}}

\newcommand{\Rom}[1]{\expandafter\@slowromancap\romannumeral #1@}
\makeatother
{ \normalfont} 

\newcommand{\m}[3]{#1_{#2 #3}}

\newcommand{\F}{\mathcal{F}}
\newcommand{\D}{\mathcal{D}}

\expandafter\def\expandafter\normalsize\expandafter{%
	\normalsize
	\setlength\abovedisplayskip{0pt}
	\setlength\belowdisplayskip{5pt}
	\setlength\abovedisplayshortskip{0pt}
	\setlength\belowdisplayshortskip{5pt}
}

\definecolor{Gray}{gray}{0.75}

\newmdenv[backgroundcolor=Gray, leftmargin = 0pt, rightmargin = 0pt, linewidth = 0pt, roundcorner = 2 pt, innerleftmargin=5pt, innerrightmargin=5pt, innertopmargin=5pt, innerbottommargin=5pt]{Frame}

\newcommand{\kk}{\langle k \rangle}
\newcommand{\kkk}{\langle k^2 \rangle}

\newcommand{\x}{{\mathbf {x}}}
\newcommand{\av}[1]{\langle #1 \rangle}

\begin{document}

\date{\today}

\begin{center}
	{\color{blue} \LARGE \textbf{Sustaining a network by controlling \\[7pt] a fraction of nodes}}
	
	\vspace{2mm}
	Hillel Sanhedrai$^{1*}$ 
	\& Shlomo Havlin$^{1}$ 
\end{center}

\small{
	\begin{enumerate}
		\item
		\textit{Department of Physics, Bar-Ilan University, Ramat-Gan, Israel}				
	\end{enumerate}

\begin{itemize}
	\item[\textbf{*}]
	\textbf{Correspondence}:\ \textit{hillel.sanhedrai@gmail.com}
\end{itemize} 

\vspace{1mm}

\textbf{
	Multi-stability is a widely observed phenomenon in real complex networked systems, such as technological infrastructures, ecological systems, gene regulation, transportation and more. When a system functions normally but there exists also a potential state with abnormal low activity, although the system is at equilibrium it might make a transition into the low activity undesired state due to external disturbances and perturbations. Thus, such a system can be regarded as \emph{unsustainable}, due to the danger of falling into the potential inactive state.
	Here we explore, analytically and by simulations, how supporting the activity of a fraction $\rho$ of nodes can turn an unsustainable system to be \emph{sustainable} by eliminating the inactive potential stable state. 
	We thus unveil a new sustainability phase diagram in the presence of a fraction of controlled nodes $\rho$. This phase diagram could provide guidelines to sustain a network by external intervention and/or by strengthening the connectivity of the network.
}	
\vspace{5mm}

\pagenumbering{arabic}


Biological, social or technological complex systems
experience in certain cases catastrophic failure causing the collapse of the whole system functionality. For instance, overload failures in power systems \cite{Yang2017,Zhao2016}, 
species extinction in ecological networks  \cite{courchamp2006rarity,shih2016ecological,jiang2019harnessing},
traffic jams in a city 
\cite{zeng2020multiple},	
and cell death in cellular dynamics  \cite{Alon2006,Jeong2001}. 
Such collapses can be caused by structural damages, causing the networks to lose their connectivity \cite{cohen-prl2000,Albert2002,watts-nature1998,barthelemy-physicsreports2011,Buldyrev2010,gao2012networks}, however systems may lose
their functionality despite they are still connected due to a sparse connectivity and/or functional disturbances \cite{Motter2002,Crucitti2004,Boccaletti2006,Dobson2007,Achlioptas2009,menck2013basin,majdandzic2014spontaneous,zhang2015explosive,Boccaletti2016,Gao2016, behar2016fluctuations, Danziger2019,sanhedrai2022reviving}.
Once a system collapses, the question is how it responds in this situation. Two types of systems can be distinguished. Some systems recover by their own and go back to normal function, while others remain in their abnormal nonactive state, and can be recovered \emph{only} by external recovery. The former kind is considered as \emph{sustainable} systems since they are permanently active even after disturbances. The latter type in contrary, is considered as \emph{unsustainable} systems, since even though they are stable, yet in the presence of external perturbations they collapse and do not return spontaneously to their original state.
In fact, what makes the difference between these two types, is the bi-stability character of the dynamics \cite{menck2013basin}. If there is another stable state, non-functional, then once the system reaches this state it stays in this state. However, when the active state is the only stable state, there is no danger to fall into another stable state since there is no such one.

In this manuscript, we aim to explore the question of transforming an \emph{unsustainable} system to a \emph{sustainable} system, by supporting the activity of a fraction of nodes. This intervention is done by forcing a fraction $\rho$ of controlled nodes to have a high value of activity $\Delta$.
We develop a framework to predict for a given network structure and for a given intervention (ruled by $\rho$ and $\Delta$) if the system is sustainable. By defining a parameter $\beta$ \cite{Gao2016}, that captures the connectivity of the network, we construct and present a new phase diagram in the ($\beta,\rho$) space.
From this phase diagram, for a given network with certain $\beta$, we can determine what is the minimal (critical) fraction of nodes $\rho_c$ that is needed for sustaining the network.
In addition, an earlier research \cite{sanhedrai2022reviving} has shown that under certain conditions even a \emph{single} node can revive a system, and therefore under these conditions of course this single node makes the system sustainable. We show here that by controlling a fraction of nodes, we enlarge considerably the sustainable phase compared to controlling a single node.
We further present a theory that bridges between the two limits of macroscopic and microscopic sets of forced nodes, covering both extremes.
We demonstrate and apply our framework on three dynamic processes, cellular, neuronal and spins dynamics, showing its generality. However, different systems show remarkable varying sustainability phase diagrams.

\subsection*{Unsustainable networks}

To find the conditions for which a network is unsustainable, we first analyze the dynamics of a free system without external intervention. 
We rely upon a general framework \cite{Barzel2013a,Harush2017,Hens2019} to model nonlinear dynamics on networks.
Consider a system consisting of $N$ components (nodes) whose activities $x_i$ ($i=1,2,...,N$) follow the Barzel-Barab\'asi \cite{Barzel2013a} equation,
\begin{equation}
\dod{x_i}{t}= M_0(x_i)+ \lambda\sum\limits_{j=1}^{N}A_{ij}M_1(x_i)M_2(x_j).
\label{Free}
\end{equation}
The first function, $M_0(x_i)$, captures node $i$'s self-dynamics, describing mechanisms such as protein degradation \cite{Barzel2011} (cellular), individual recovery \cite{Dodds2005,PastorSatorras2015} (epidemic) or birth/death processes \cite{Gardner2000} (population dynamics). The product $M_1(x_i)M_2(x_j)$ describes the $i,j$ interaction mechanism, representing \textit{e.g.}, genetic activation \cite{Alon2006,Karlebach2008,schreier2017exploratory}, infection \cite{Dodds2005,PastorSatorras2015} or symbiosis \cite{Holling1959}. The binary adjacency matrix $A$ captures the network, \textit{i.e.}\ the interactions (links) between the nodes. An element $\m Aij$ equals $1$ if there is an interaction (link) between nodes $i$ and $j$ and $0$ otherwise. A is symmetric and obeys the configuration model characteristics. The strength of the interactions is governed by the positive parameter $\lambda$. 

In Fig.\ \ref{fig:Illustration} we demonstrate our problem on the example of gene regulation dynamics.
For a weak connectivity, expressed by low interactions strength or small density of links, there exists only the low-active state where all genes are suppressed, Figs.\ \ref{fig:Illustration}a,b, while for a strong connectivity, there emerges an additional high-active state. However, the low-active state still exists, Figs.\ \ref{fig:Illustration}d,e, and allows the risk of collapsing from functionality into non-functionality as a result of some disturbances, see Fig.\ \ref{fig:Illustration}f. Therefore, such active state is \emph{unsustainable} due to the potential failure into the inactive state. In this study, we show how supporting a small fraction of the system nodes (Figs.\ \ref{fig:Illustration}g,h, dark blue) \emph{eliminates} the nonfunctional state and as such, makes the system sustainable to perturbations, Fig.\ \ref{fig:Illustration}i.

\begin{figure}[h!]
	\centering
	\includegraphics[width=0.95\linewidth]{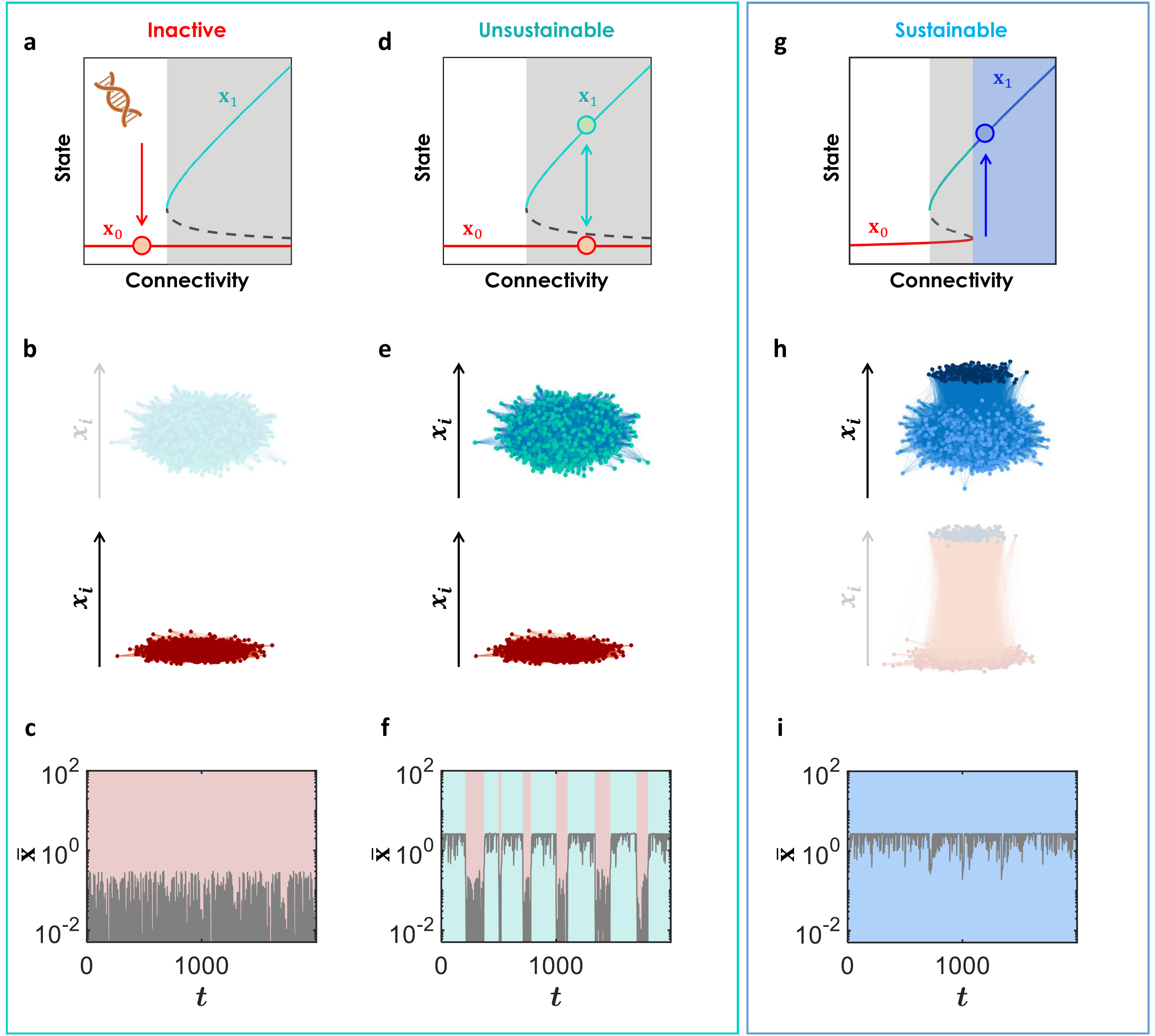}
	\caption{ 
		\textbf{ 
		The challenge of unsustainable network and how to make it sustainable.} 
		(a) Diagram of regulatory dynamics showing two states, inactive $\x_0$ and active $\x_1$. For a weak connectivity only $\x_0$ exists, thus the system resides in the inactive phase.		
		(b) For a weak connectivity, there is no active stable state (green, very light), but just the inactive state (red).
		The vertical ($z$) coordinate represents the activity of each node.
		(c) Adding random perturbations (noise) to the activities does not help to activate the network since a functional stable state does not exist, thus the average activity stays low all the time.
		(d) For a dense connectivity both states are stable, such that there exists a bi-stable region (gray shade). 
		(e) A system located in the bi-stable regime, demonstrates two stable states, active (green) and suppressed (red). Thus, $\x_1$ is \emph{unsustainable} in this region since the inactive state $\x_0$ also exists. 		 
		(f) An implication of the bi-stability is that adding random perturbations (noise) to the activities, results in spontaneous system transitions between the active (green) and inactive states (red) since both are stable in the \emph{unsustainable} phase. 
		(g)-(h) Here we consider the same system as in (d), but we control a fraction of the system (dark blue circles), $\rho$, and hold it with a constant high value, $\Delta$. Due to this intervention, the low-functional state vanishes what makes the system \emph{sustainable}.		
		(i) When we control a fraction of nodes with high activity, perturbations such as those in (f) are not capable to deflect the system from the active state, since the inactive state disappears. Namely, the control makes the system \emph{sustainable}.
	}
	\label{fig:Illustration}
\end{figure}

\subsection*{Sustaining a network} 

To drive an unsustainable network to be sustainable we consider a simple intervention. We \emph{force} a set of nodes $\F$ (fraction $\rho$) to have a constant high activity value $\Delta$ (Fig.\ \ref{fig:Illustration}h), while all the rest in the complementary set $\D$ are governed by the original dynamics. 
Thus, such a forced system obeys the set of equations,
\begin{equation}
	\left\{
	\begin{array}{cclr}
		x_i &=& \Delta & i \in \mathcal{F}
		\\[7pt]
		\dod{x_i}{t} &=& M_0(x_i) + \lambda \displaystyle \sum_{j = 1}^N \m Aij M_1(x_i) M_2(x_j) \hspace{7mm} & i \in \D 
	\end{array}
	\right. .
	\label{ForcedDynamics}
\end{equation} 
Next we aim to track the states of the unforced nodes, \textit{i.e.}\ the set $\D$.
For finding the steady states of the forced system, we demand a relaxation, thus the derivative vanishes. In addition, we use a Degree-Based Mean-Field approximation \cite{PastorSatorras2001prl,boguna2002epidemic,Barrat2008,Dorogovtsev2008,PastorSatorras2015} assuming nodes having the same degree behave similarly. Thus, we replace the binary term $\m Aij$ with the probability of link existence between $i$ and $j$ in the configuration model, $k_ik_j/(N\av{k})$ where $k_i$ and $k_j$ are the degrees of $i$ and $j$ respectively. Considering this, we define (see elaboration in SI Section 2)
the order parameter $\Theta$ as
\begin{equation} \label{eq:Theta}
	\Theta =  \frac{1}{|\D|\av{k_{\D\to\D}}} \sum_{j \in \D} k_j^{\D\to\D}  M_2(x_j^*) ,
\end{equation}
which represents the mean impact that an arbitrary free node gets from its arbitrary free neighbor. The variable $x_j^*$ stands for the activity of node $j$ in relaxation, and $\D\to\D$ means to count only links within $\D$, \textit{i.e.}\ the unforced nodes.
Using the defined $\Theta$, and applying the mean-field approximation in Eq.\ \eqref{ForcedDynamics} in relaxation, we obtain,
\begin{equation}
	R(x_i^*) = \lambda k_i^{\D\to\D} \Theta  + \lambda k_{i}^{\D\to\F}  M_2(\Delta),
	\label{RxiTheta}
\end{equation}
where $R(x)=-M_0(x)/M_1(x)$, and $k_i^{\D\to\F}$ is the number of forced neighbors (in $\F$) of node $i$ which is a free dynamic node (in $\D$).
Substituting this in the definition of $\Theta$, we obtain a self-consistent equation for the order parameter,
\begin{equation}
	\Theta =  \frac{1}{|\D|\av{k_{\D\to\D}}}  \sum_{j \in \D} k_j^{\D\to\D}  M_2(R^{-1}(\lambda k_j^{\D\to\D} \Theta  + \lambda k_{j}^{\D\to\F}  M_2(\Delta))) ,
	\label{eq:TheteSC}
\end{equation}
where $R^{-1}$ is the inverse function of $R$. This step assumes that $R$ is an invertible function.
Solving the self-consistent Eq.\ \eqref{eq:TheteSC}, we get all the states (both stable and unstable) of the system. This equation can be solved using any degree distribution and the specific selection of the nodes in $\F$ (see SI Section 2.2 and Fig.\ \ref{fig:Theta}).

To obtain more intuitive, simple and useful expressions, we assume the following. When the degree distribution is not very broad and/or the functions $M_2$ and $R$ are close to linear or constant, we insert the average into the functions \cite{Gao2016}, \textit{i.e.}\ $\overline{M_2(\x)}  = M_2(\bar{\x})$ and $\overline{R(\x)} = R(\bar{\x})$, see also SI Section 2.2.2. This allows us, using Eqs.\ \eqref{eq:Theta} and \eqref{RxiTheta}, to obtain a very simple relation between the average steady state $\bar{\x}$ of the free nodes and the connectivity $\beta$ of the network.
We define the average activity $\bar{\x}$ over all the neighbors within $\D$ by
\begin{equation} \label{eq:xBar}
	\bar{\x} = \frac{1}{|\D|\av{k_{\D\to\D}}} \sum_{j \in \D} k_j^{\D\to\D}  x_j^* ,
\end{equation}
and the connectivity $\beta$ is defined as
\begin{equation}
	\beta = \lambda \kappa,
	\label{Beta}
\end{equation}
combining both the interactions strength $\lambda$ and the average neighbor degree over the whole network, $\kappa = \kkk/\kk$. Using these terms, we finally obtain a simple equation for the states of a forced system, for $\kappa\gg1$ and random selection of the forced nodes,
\begin{equation}
	\beta = \frac{ R(\bar{\x}) }{(1-\rho) M_2(\bar{\x}) + \rho  M_2(\Delta)},
	\label{betaXForced}
\end{equation}
where $\rho=|\F|/N$ is the fraction of the controlled nodes. To get the free system states, we can just substitute $\rho=0$, yielding $\beta=R(\bar{\x})/M_2(\bar{\x})$. (For non-random selection of the forced nodes, see SI Section 3.)

Eq.\ \eqref{betaXForced} implies that forcing a fraction $\rho$ of nodes to have an $x_i$-value $\Delta$ changes the phase diagram of the system, and creates a new phase diagram for a forced system. To demonstrate this change in the phase diagram, we go back to our main example of gene regulation. The free system, Fig.\ \ref{fig:Theory}b top, exhibits two regimes: inactive state for weak connectivity, and bi-stable regime (gray shade) above a certain $\beta$. In marked contrast, a forced system shows a remarkable different phase diagram, Fig.\ \ref{fig:Theory}b bottom, exhibiting three regimes: inactive for small $\beta$, bi-stability for intermediate $\beta$, and above certain value of $\beta$ only an active state. 
Neuronal and spins dynamics, Figs.\ \ref{fig:Theory}c,d, have distinct phase diagrams which change as well by controlling a fraction of the system. 
In all these three examples, a free system located at the high-active state within the bi-stable area (gray shade), is regarded \emph{unsustainable}. This is since there exists a potential inactive state. However, controlling a fraction $\rho$ of nodes with high activity $\Delta$, reshapes the phase diagram, and creates an area (blue shade), in which the system becomes safer, that is with no risk of failure into the low state. Hence the system becomes \emph{sustainable}.

\begin{figure}[ht]
	\centering
	\includegraphics[width=0.95\linewidth]{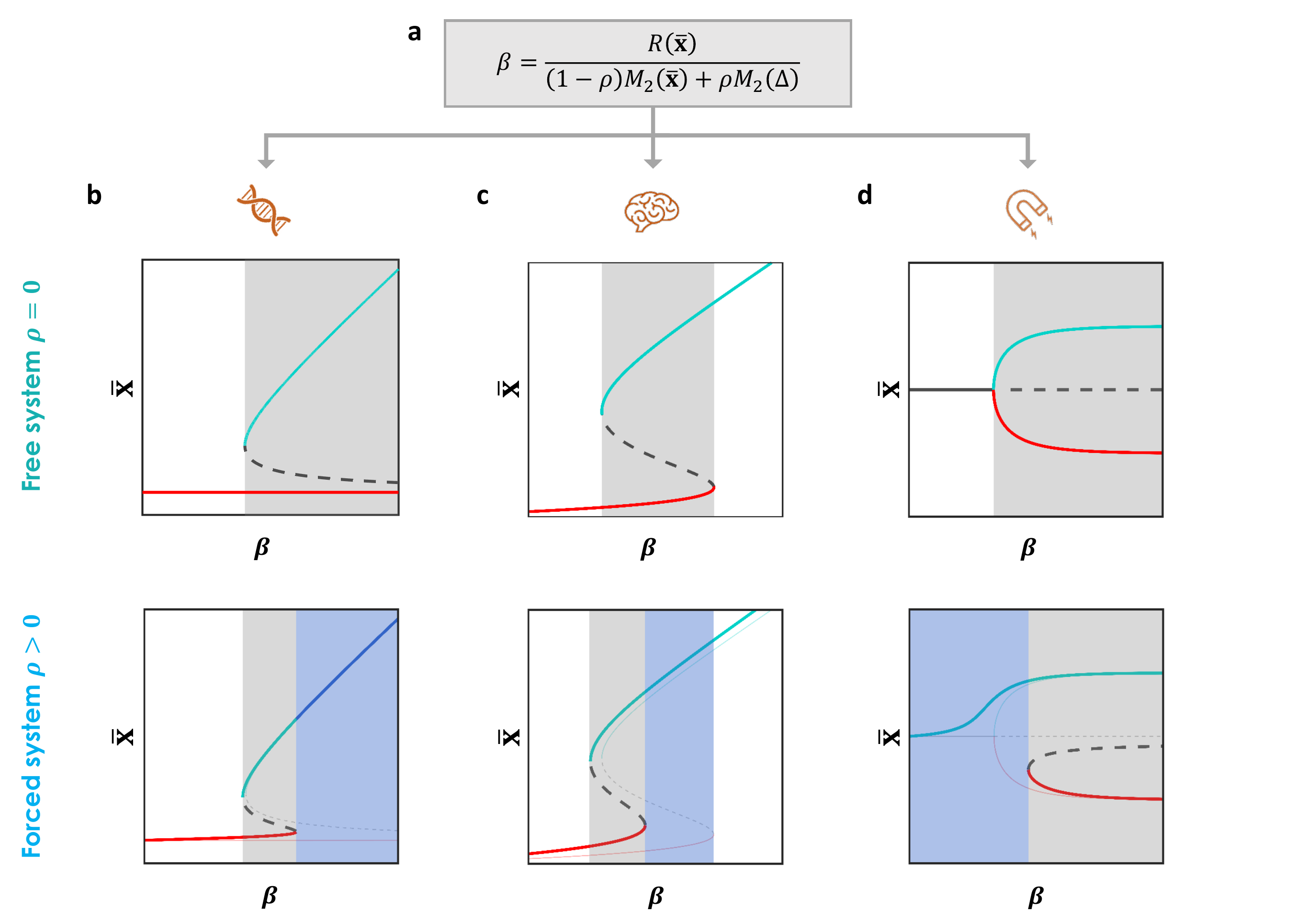}
	\caption{ 
		\textbf{Theory. Forced system has a new phase diagram.} 
		(a) The relation, Eq.\ \eqref{betaXForced}, between the system state ($\av{\x}$) and its connectivity ($\beta$) provides the phase diagrams of both free ($\rho=0$) and controlled ($\rho>0$) system for any dynamics. Here we demonstrate this for three distinct systems. 
		(b) Cellular dynamics (see Fig.\ \ref{fig:Cellular}). A free system diagram (top) shows a suppressed function at $\x_0$ for small $\beta$, and a bi-stable regime for large $\beta$, where $\x_0$ and $\x_1$ both exist and are stable. Thus the system is unsustainable and has a risk to fail into the non-functional state. 
		In contrast, a forced system (bottom), controlled by holding a fraction $\rho$ of nodes with high value $\Delta$ shows a new phase diagram, having a s-shape curve, with a new region for large $\beta$ in which only $\x_1$ appears (blue).
		Thus, a system in the blue region is now safe and not only active, but also sustainable.		
		(c) Brain dynamics (see Fig.\ \ref{fig:BrainSpin}a-d) exhibits three regimes:\ inactive for sparse topology, active for dense topology and bi-stable in between.
		Forcing the system (bottom) with certain $\Delta$ and $\rho$ pushes the twist of the s-shape to a lower $\beta$-value, and as such makes the blue regime becoming sustainable rather than bi-stable.		
		(d) Spin dynamics (see Fig.\ \ref{fig:BrainSpin}e-h) has a zero inactive stable state for a sparse connectivity, while for dense connectivity there are two symmetric active states. Controlled system shows a new phase diagram including a sustainable regime where only the positive stable state appears (blue shade). This dynamics does not fall into the formula in (a), however a similar analysis can be done, see SI Section 5.3. 
	}
	\label{fig:Theory}
\end{figure}

\subsection*{Application:\ Cellular dynamics} 

As our main example in this paper, we apply our framework on the regulatory dynamics, captured according to Michaelis-Menten model \cite{Karlebach2008}, by
\begin{equation}
	\dod{x_i}{t}= -Bx_i^a+ \lambda\sum\limits_{j=1}^{N}A_{ij} \frac{x_j^h}{1+x_j^h}.
	\label{Regulatory}
\end{equation}
Under this framework $M_0(x_i) = -B x_i^a$, describing degradation ($a = 1$), dimerization ($a = 2$) or a more complex bio-chemical depletion process (fractional $a$), occurring at a rate $B$; without loss of generality we set here $B = 1$. The activation interaction is captured by the Hill function of the form $M_1(x_i) = 1$, $M_2(x_j) = x_j^h/(1+x_j^h)$, a \textit{switch-like} function that saturates to $M_2(x_j) \rightarrow 1$ for large $x_j$, representing $j$'s positive, albeit bounded, contribution to node $i$ activity, $x_i(t)$. 

When analyzing this system while it is forced by a fraction $\rho$ of random nodes with activity $\Delta$, we obtain, using Eq.\ \eqref{betaXForced},
\begin{equation}
	\beta = \frac{\bar{\x}^a}{(1-\rho)/(1+\bar{\x}^{-h})+\rho/(1+\Delta^{-h})}.
	\label{betaXRegulatory}
\end{equation}

In Fig.\ \ref{fig:Cellular} we present in detail the results for cellular dynamics using simulations and theory when setting $a=1$, $h=2$.
The phase diagram of a free system, derived from Eq.\ \eqref{betaXRegulatory} by substituting $\rho=0$, is shown in Fig.\ \ref{fig:Cellular}b. As explained above, the high-active state $\x_1$ is unsustainable for the full range as demonstrated in Fig.\ \ref{fig:Cellular}c. 
Eq.\ \eqref{betaXRegulatory} generates also the phase diagram for a forced system shown in Fig.\ \ref{fig:Cellular}d (thick curve) for $\rho=0.03$ and $\Delta=5$, exhibiting an s-shape diagram which has now also a new only-active regime (blue shade). This regime is the \emph{sustainable phase}. In Fig.\ \ref{fig:Cellular}f we demonstrate the forced system states in the three distinct regimes, and show that for $\beta=3.1$ the external intervention makes the system sustainable rather than unsustainable in Fig.\ \ref{fig:Cellular}c. In Figs.\ \ref{fig:Illustration}f,i we demonstrate the implication of the low state disappearance to random perturbations.
The s-shape unveils a critical value of $\beta=\beta_c$ above which the system is sustainable. This $\beta_c$ depends on $\rho$, and this relation holds in the inverse direction as well, namely for given $\beta$ there is a required critical fraction $\rho_c$ of controlled nodes to make a system sustainable.
Therefore, we move to find the relation between the critical values of $\rho$ and $\beta$ at the transitions from the bi-stable region to both the sustainable phase, as well as to the inactive region. Both transitions are local extremum of $\beta(\bar{\x})$, and hence they are found using Eq.\ \eqref{betaXForced} by
\begin{equation} \label{eq:maximum}
	\frac{\partial \beta}{\partial \bar{\x}}\bigg|_{\beta_c} = 0 .
\end{equation}
Eq.\ \eqref{eq:maximum} provides $\beta_c$ as a function of both, the fraction $\rho$, and the force of control $\Delta$. Using Eq.\ \eqref{betaXRegulatory} we get $\beta_c$ or $\rho_c$ for cellular dynamics. For gene regulation, the solution is not trivial (see SI Section 5.1.3), however, in the limit of small $\rho$, when we get a small value of $\bar{\x}$ at the transition as in Fig.\ \ref{fig:Cellular}d, we obtain the scaling relation,
\begin{gather} \label{eq:rho}
	\rho_c \sim \beta ^{h/(a-h)}.
\end{gather}
Indeed, in our simulations, for the values, $a=1,h=2$, we obtain, in Fig.\ \ref{fig:Cellular}k, $\rho_c\sim\beta^{-2}$, where $\beta=\lambda\kappa$. We discuss this result further below.

Fig.\ \ref{fig:Cellular}e shows similar results as \ref{fig:Cellular}d for a larger fraction of controlled nodes, $\rho=0.11$. One can see that for this value of $\rho$, the bi-stable area almost completely disappears. This suggests another interesting critical fraction $\rho_0$, above which the active state of the system becomes sustainable for any $\beta$. To find this $\rho_0$, we notice that it captures a merge of local maximum and minimum, thus beside Eq.\ \eqref{eq:maximum}, also the second derivative should vanish, 
\begin{equation} \label{eq:maximum2}
	\frac{\partial^2 \beta}{\partial \bar{\x}^2}\bigg|_{\beta_c} = 0.
\end{equation}
These two conditions, Eqs.\ \eqref{eq:maximum} and \eqref{eq:maximum2}, together determine a tricritical point $(\beta_0,\rho_0)$ in the $(\beta,\rho)$-space at which the three phases:\ inactive, unsustainable and sustainable meet, and beyond which the system will not experience an abrupt transition at all, see Fig.\ \ref{fig:Cellular}g. For the cellular dynamics we find that (see SI Section 5.1.4),
\begin{equation} \label{eq:rho0}
	\rho_0 = \bigg(1+\frac{4ah}{(1+\Delta^{-h})(h-a)^2}\bigg)^{-1},
\end{equation}
which for our values $a=1$, $h=2$, and for a large $\Delta$, is equal approximately to $1/9$, while using $\Delta=1$ it gets an higher value of $1/5$, see Fig.\ \ref{fig:Cellular}g,h.

In Fig.\ \ref{fig:Cellular}g,h we present the new sustainability phase diagram in $(\beta,\rho)$-space for networks with $\kappa=20,60,100$ showing a good agreement between simulations and theory, Eqs.\ \eqref{betaXRegulatory} and \eqref{eq:maximum}. Note that, as expected, for larger $\kappa$, the theory agrees better with the simulations. Furthermore, one can see that the tricritical point is in agreement with Eq.\ \eqref{eq:rho0}, and gets higher for smaller $\Delta$, Fig.\ \ref{fig:Cellular}h. The symbols in Fig.\ \ref{fig:Cellular}g refer to the demonstrations in Fig.\ \ref{fig:Cellular}f exhibiting the system state in each region.

\begin{figure}[h]
	\centering
	\includegraphics[width=0.95\linewidth]{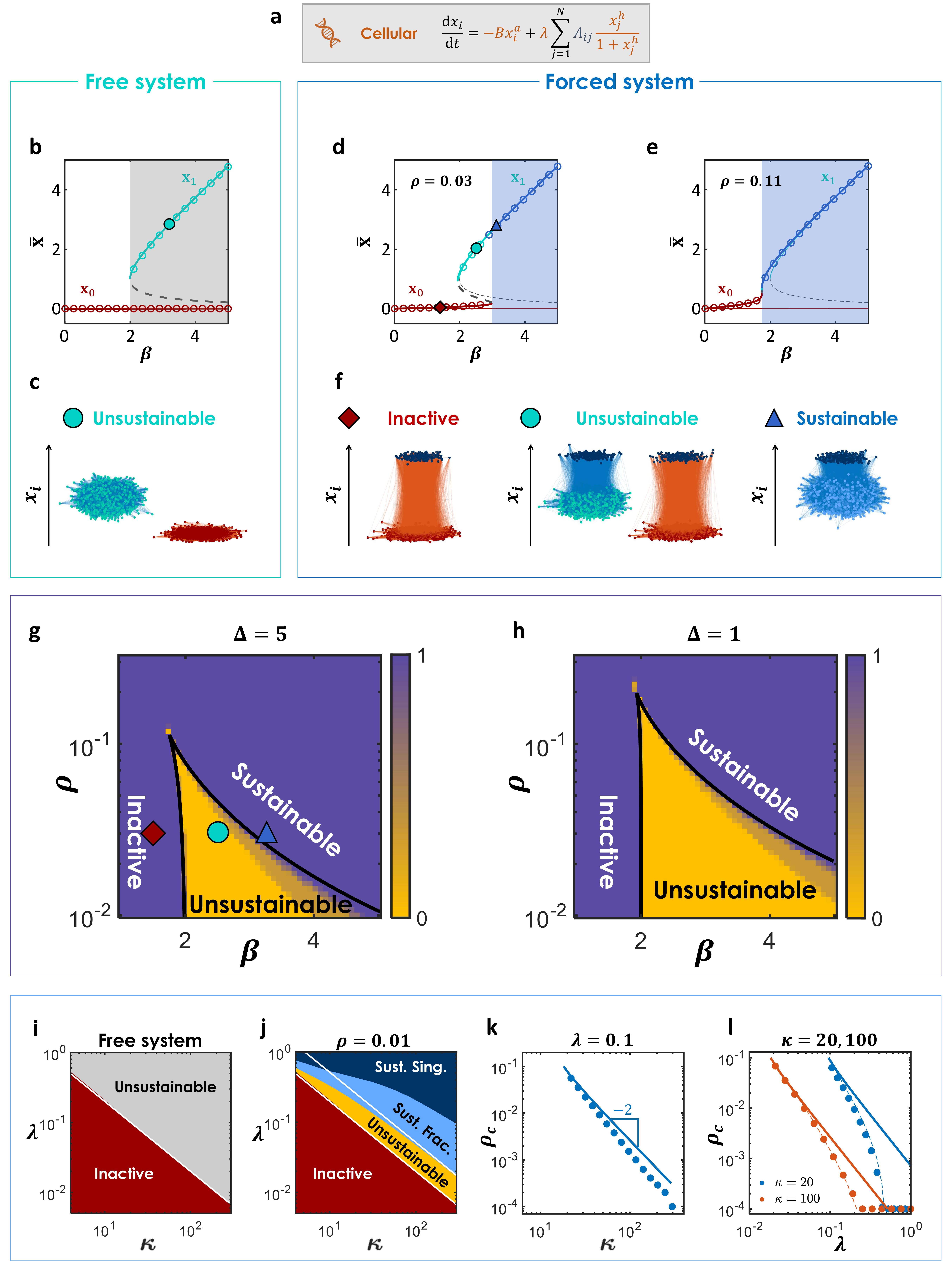}
\end{figure}

\clearpage 
\captionof{figure}{	
	\textbf{Sustainability of a cellular network.} 
	(a) We apply our framework on the regulatory dynamics captured by Michaelis-Menten model \cite{Karlebach2008}. 
	(b) Simulations (symbols) and theory (lines, Eq.\ \eqref{betaXRegulatory} with $\rho=0$) results for a free system of ER structure with $N=10^4$ and $\kappa=40$. There is a bi-stable region (gray shade).
	(c) Demonstration of the activities of the system with $\beta=3.1$ for $\rho=0$. Both states are stable, therefore the system is unsustainable.
	(d) For a forced system by a fraction $\rho=0.03$ of random nodes with activity $\Delta=5$, Eq.\ \eqref{betaXForced} provides a phase diagram (thick curve), exhibiting an s-shape curve which has now also a regime with only a single active state (blue shade). This regime is a sustainable phase. Note that simulations (symbols) are in agreemant with the theory (lines). The network is the same as in (b).
	(e) The same as (d) with a larger fraction of controlled nodes, $\rho=0.11$. Here we see that the unsustainable region almost vanishes, and the transition becomes almost continuous. This agrees with Eq.\ \eqref{eq:rho0}.
	(f) Activities for $\beta=1.5,2.5,3.1$ in three regions, inactive (red diamond), unsustainable (green circle) and sustainable (blue triangle) correspondingly. The dark blue nodes are the forced nodes. The red nodes represent $\x_0$, the green nodes represent unsustainable $\x_1$, and the light blue nodes represent sustainable $\x_1$.
	(g) The new phase diagram in $(\beta,\rho)$-space for $\Delta=5$. The simulations were done on ER networks with $N=10^4$ for 50 values of $\rho$, 50 values of $\beta$, for $\kappa=20,60,100$, and averaged over 10 realizations. In the color bar, the value 0 represents an unsustainable system, and 1 represents the other cases. The black lines are obtained from Eqs.\ \eqref{betaXRegulatory} and \eqref{eq:maximum}.
	(h) The same as (g) with lower intervention force $\Delta=1$. As expected in this case a larger fraction of controlled nodes is needed to make the network sustainable.
	(i) $(\kappa,\lambda)$-space for a free system.	
	(j) The sustainability phase diagram for  $\rho=0.01$ and $\Delta=5$. The light blue is the sustainable phase when forcing a fraction $\rho=0.01$ of nodes, and the dark blue is the sustainable regime for holding a single node. 
	The white lines represent the theory, Eqs.\ \eqref{betaXRegulatory} and \eqref{eq:maximum}.	
	(k) Horizontal trajectory in $(\kappa,\lambda)$-space for fixed $\lambda=0.1$ and varying $\kappa$. Symbols are simulations and the line is theory obtained from Eqs.\ \eqref{betaXRegulatory} and \eqref{eq:maximum}. The slope is according to Eq.\ \eqref{eq:rho}.
	(l) Vertical trajectory in ($\kappa,\lambda$)-space for fixed $\kappa=20,100$ and varying $\lambda$. Note that the critical fraction for sustaining, $\rho_c$, for a given $\lambda$ approaches 0 where it reaches the single-node sustainable phase in simulations (symbols). The theory, Eqs.\ \eqref{betaXRegulatory} and \eqref{eq:maximum} (continuous lines), deviates from the simulations results for small $\rho$. The dashed lines are from a different theory, see SI Section 4, which captures also the limit of small $\rho$.	
} 
\label{fig:Cellular}	

\rule{\textwidth}{0.3mm}
\vspace{3mm}

Next, we split the merged connectivity $\beta$ and move to look on the 3D space $(\kappa,\lambda,\rho)$, and particularly on the $(\kappa,\lambda)$-space. This is since we want to test the system behavior also for low degrees, and thus we decouple $\beta$ into two parameters $\kappa$ and $\lambda$, Eq.\ \eqref{Beta}. Moreover, we want to generalize the phase diagram to include the single-node reviving model \cite{sanhedrai2022reviving}, where under certain conditions controlling one node revives the whole system into its high-active state. When it works, it also makes the system sustainable since it cancels the low-active state. 
Fig.\ \ref{fig:Cellular}i shows the phase diagram of a free system with two areas as above in Fig.\ \ref{fig:Cellular}b. In Fig.\ \ref{fig:Cellular}j we show the combined single-node reviving (dark blue), and fraction-reviving with $\rho=0.01$ (light blue). The control of a small fraction of the system considerably extends the sustainable phase. The white lines represent the results of our theory, calculated via $\beta_c$ using Eqs.\ \eqref{betaXRegulatory} and \eqref{eq:maximum}. Note that the theory works well for high $\kappa$ while deviates for small values. The reason is the nature of the mean field approximation which works usually quantitatively only for large degree. The reason is that if nodes have more neighbors, then different neighborhoods are more similar, as assumed by mean-field theory.

In Fig.\ \ref{fig:Cellular}k we cross the $(\kappa,\lambda)$-space horizontally for fixed $\lambda=0.1$ while changing $\kappa$. Symbols represent simulations and line represents theory, obtained by Eqs.\ \eqref{betaXRegulatory} and \eqref{eq:maximum}. This supports the scaling obtained in Eq.\ \eqref{eq:rho} $\rho_c\sim\beta^{-2}$. See SI Fig.\ S3 for other values of $a$ and $h$. It can be seen that for very small $\rho$ there is an increasing deviation even though $\kappa$ is large. This slight discrepancy, becomes larger in Fig.\ \ref{fig:Cellular}l where we move vertically with $\lambda$ for fixed $\kappa=20$ and $\kappa=100$. Here the deviation of the theory (continuous line) for small $\rho$ is significant. However it gets better for larger degrees as expected. This gap exposes the contradiction between our MF and simulations regarding the limit of $\rho\to0$, as discussed below.

\subsubsection*{The limit \boldmath$\rho\to0$ }

Our MF theory, by definition yields that $\rho\to0$ derives $\beta_c\to\infty$ as in a free system without intervention at all. However, it has been shown \cite{sanhedrai2022reviving} that for large enough $\lambda$ even a single node impacts the network globally. The reason that our MF fails for very small $\rho$ and not very large degree is that when there are only one or few controlled nodes the uniformity or homogeneity assumption of different nodes in the network, upon which the MF relies, is broken. The network gets a structure of shells around the few controlled nodes as analyzed in \cite{sanhedrai2022reviving}. 
Therefore, we developed here a modified shells MF which includes also the case of many controlled nodes rather than only one. This method bridges between the microscopic and the macroscopic intervention limits. See SI Section 4 for details. The dashed lines in Fig.\ \ref{fig:Cellular}l are obtained from this improved method, showing excellent agreement with simulations, thus, covering both extremes.

\subsubsection*{Large fluctuations}

Another challenge for our MF theory is large fluctuations. This is  when the network has a small average degree, but particularly for scale-free networks which exhibit a broad degree distribution, which represent large variations in node degrees. For these cases our MF approach does not work. To overcome the challenge of such networks, we step back in our theory derivations to Eq.\ \eqref{eq:TheteSC}, and use the order parameter $\Theta$, Eq.\ \eqref{eq:Theta}, to determine the system states. This method does not assume the general parameter $\beta=\lambda\kappa$, and even not the global parameter $\kappa$. But the analysis and solution depend on $\lambda$ and on the \emph{full degree-distribution} $p_k$, Eq.\ \eqref{eq:TheteSC}. In Fig.\ \ref{fig:Theta} we present the advantage of this method compared to the above method, represented by Eq.\ \eqref{betaXForced}. Fig.\ \ref{fig:Theta}a shows the very good agreement between simulations and theory, Eq.\ \eqref{eq:TheteSC} for scale-free network with $\gamma=3.5$ and $k_0=15$. In Fig.\ \ref{fig:Theta}b we present the ($\beta,\rho$) phase diagram for the same scale-free network as in Fig.\ \ref{fig:Theta}a. The continuous lines represent the theory of Eq.\ \eqref{eq:TheteSC} showing excellent agreement with the simulations results, while the dashed lines representing Eq.\ \eqref{betaXForced} fail. The SF's phase diagram implies the higher susceptibility of SF to become sustainable compared to ER. For example, as seen in Figs.\ \ref{fig:Theta}c and \ref{fig:Theta}b, $\rho_0$ that ensures the absence of unsustainability is about $0.1$ in ER while it is only about $0.03$ in the presented SF. SF with smaller $\gamma$, and smaller $k_0$, challenges also Eq.\ \eqref{eq:TheteSC}, see SI Fig.\ S4. 
Finally, we show in Fig.\ \ref{fig:Theta}c that also in ER with not very large degree, $\kappa=10$, Eq.\ \eqref{eq:TheteSC} supplies a significant better accuracy (full line) compared to Eq.\ \eqref{betaXForced} (dashed line).

\begin{figure}[h]
	\centering
	\includegraphics[width=0.99\linewidth]{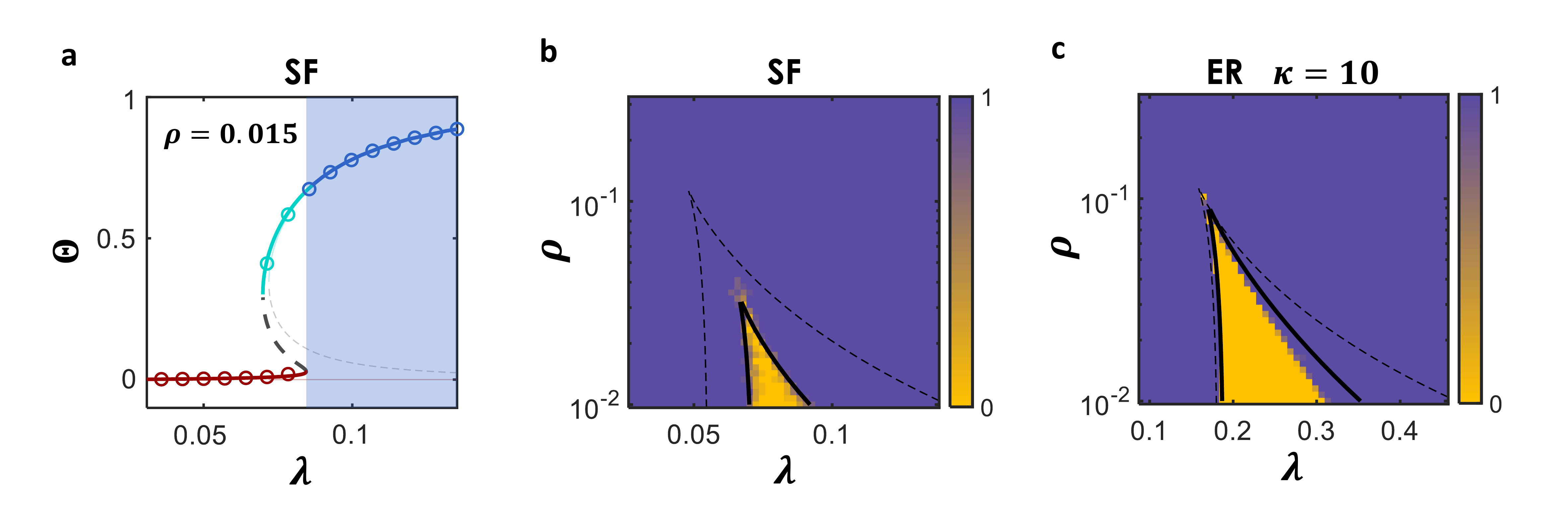}
	\caption{ \textbf{Large fluctuations.}
		Results for scale-free networks and ER with low degrees which have large fluctuations, and thus demand analysis using Eq.\ \eqref{eq:TheteSC} rather than Eq.\ \eqref{betaXForced}.
		(a) The order parameter $\Theta$, Eq.\ \eqref{eq:Theta}, for a forced system with $\rho=0.015$ and $\Delta=5$. The network here is scale-free with $N=10^4$, minimal degree $k_0=15$, and exponent $\gamma=3.5$. The symbols are from simulations and the line is from theory, Eq.\ \eqref{eq:TheteSC}.
		(b) For the same network as in (a), the $(\lambda,\rho)$ phase diagram shows significantly narrower unsustainable regime compared to ER. The black continuous lines are from Eq.\ \eqref{eq:TheteSC} while the dashed lines are from Eq.\ \eqref{betaXForced} which captures well ER networks but fails to predict SF.
		(c) Results for ER network with $\kappa=10$ which is smaller degree than in Fig. \ref{fig:Cellular}g. Also here the theory of Eq.\ \eqref{eq:TheteSC} (full line) is better than Eq.\ \eqref{betaXForced} (dashed line).
	}
	\label{fig:Theta}
\end{figure}

\subsection*{Additional examples}

Next, we consider other two types of dynamics to exhibit aspects that do not appear in the cellular dynamics as well as to demonstrate the generality of our framework.

\subsubsection*{Neuronal dynamics}

As our second example we consider neuronal dynamics governed by the set of equations given in Fig.\ \ref{fig:BrainSpin}a, based on a modification of Wilson-Cowan model \cite{wilson1972excitatory,wilson1973mathematical}, see SI Section 5.2. As can be seen in Fig.\ \ref{fig:BrainSpin}b (thin light lines) the system naturally exhibits an s-shape curve including three dynamic phases. The \textit{inactive} phase for small $\beta$ where there exists only the state $\x_0$, in which all activities are suppressed.
The region of high $\beta$, where there exists only $\x_1$ in which the activities $x_i$ are relatively high. Thus, for high $\beta$ the system is \emph{naturally sustainable}. In between these two extremes, the system features a \textit{bi-stable} phase, in which both $\x_0$ and $\x_1$ are potentially stable, therefore the active state in this region is unsustainable. 
However, controlling a fraction of $\rho=0.02$ with activity of $\Delta=15$, creates a new s-shape curve with more narrow bi-stable region (thick lines and symbols). Thus, a window of sustainability (blue shade) is created, in which the control drives the system to be sustainable.
In Fig.\ \ref{fig:BrainSpin}c we observe, in $(\kappa,\lambda)$-space, for $\rho=0.02$, additionally to the three phases found for gene regulation (Fig.\ \ref{fig:Cellular}j), the naturally sustainable phase where the system is sustainable by its own without any intervention. 
Note also that our theory, Eqs.\ \eqref{betaXForced} and \eqref{eq:maximum}, predicts the required $\lambda_c$ to sustain the system given $\rho=0.02$ for high degrees (white line). 
Fig.\ \ref{fig:BrainSpin}d shows the $(\beta,\rho)$ phase diagram for $\kappa=20,60,100$.
\\

%

\subsubsection*{Spin dynamics} 

As our final example we explore the dynamics of spins connected by ferromagnetic interactions, captured by the equations of Fig.\ \ref{fig:BrainSpin}e which are based on Ising-Glauber model \cite{krapivsky2010kinetic}. 
This example is different from the two above examples since the interactions in this dynamics are \emph{attractive} rather than \emph{corroborative}. Moreover, the interactions act completely symmetrically towards both stable states in contrast to the above examples where they only push towards the high-active state. In addition, the form of equations is not included in our framework, Eq.\ \eqref{Free}, however a similar analysis holds for this dynamics, see SI Section 5.3.

Fig.\ \ref{fig:BrainSpin}f shows the significant change of the phase diagram of the free system (thin lines) due to the controlled nodes (thick lines) with a fraction $\rho=0.1$ having activity $\Delta=1$. In contrary to the symmetric states of a free system (thin and light lines), the forced system shows a region (blue shade) where $\x_1$ is sustainable. Interestingly, this area is obtained for weak connectivity (small $\beta$) differently from the above examples. This is an outcome of the \emph{attractive} nature of the interactions, since it causes a competition between the controlled nodes which pull up and the free nodes which pull down. The free nodes have a numerical advantage, however for small $\beta$ the natural negative solution is small (in absolute value), and for large enough $\Delta$, the forced nodes win. 

In Fig.\ \ref{fig:BrainSpin}g we show the phase diagram in ($\kappa,\lambda$) space as above. For spin dynamics, the phase of sustainable by a single node does not exist, due to the symmetric attractive interactions. 
The reason is that when the external signal propagates from the source node through neighbors they have on average only one neighbor behind that pulls them up and more than one neighbor that pull them down. Because of the symmetry between states, the numerical advantage wins. Thus, even if the intervention force $\Delta$ is high, its impact decays with distance and makes only a local impact in an infinite system. 
Fig.\ \ref{fig:BrainSpin}h shows the ($\beta,\rho$) phase diagram which has a different shape compared to the above examples as explained above. Note also that in this case there is no tricritical point.


\begin{figure}[ht]
	\centering
	\includegraphics[width=0.9\linewidth]{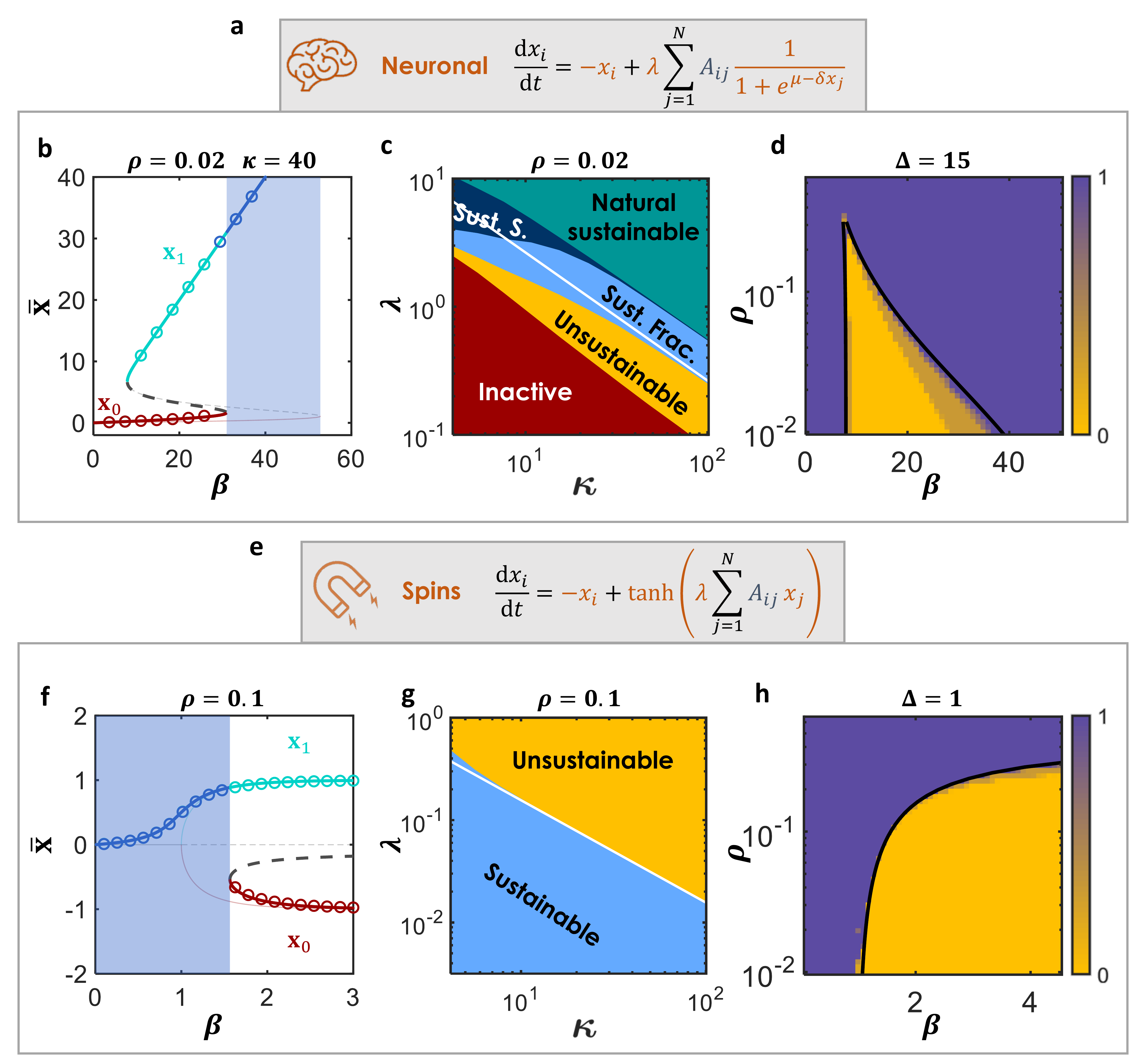}
	\caption{ \textbf{Sustainability of neuronal and spins dynamics.} 
		(a) Neuronal dynamics based on Wilson-Cowan model \cite{wilson1972excitatory,wilson1973mathematical}. 
		(b) Phase diagram for dynamics of forced system via fraction $\rho=0.02$ and holding value $\Delta=15$ according to Eq.\ \eqref{betaXForced}. The network is ER with $N=10^4$ and $\kappa=40$. The forced curve is shifted left relative to the free system curve (thin and light lines), yielding a window of sustainability (blue shade) 
		(c) $(\kappa,\lambda)$-space shows a sustainability phase diagram containing five phases. Our theory, Eqs.\ \eqref{betaXForced} and \eqref{eq:maximum}, predicts for high degrees well the transition between unsustainable and sustainable by fraction $\rho=0.02$ and $\Delta=15$ (the white line). The light blue area is the sustainable phase for controlling a fraction $\rho=0.02$, and the dark blue phase is the sustainable region when controlling a single node. 
		(d) $(\beta,\rho)$ phase diagram for $\kappa=20,60,100$. 		
		(e) Model for spin dynamics based on Ising-Glauber model \cite{krapivsky2010kinetic}. 
		(f) While free system (light and thin lines) shows a diagram with a zero regime and bi-stable symmetric regime, the forced system (thick lines) exhibits two regions:\ for a dense network (large $\beta$) coexistence of $\x_0$ and  $\x_1$, and for a sparse network (small $\beta$) only $\x_1$ appears. Consequently, there is a range of sustainability (blue shade). Here $\rho=0.1$ and $\Delta=1$. The network is ER with $N=10^4$ and $\kappa=40$. 
		(g) The $(\kappa,\lambda)$ phase diagram for $\rho=0.1$ and $\Delta=1$. Here there is no sustainable phase when holding a single node since controlling single node does not change the global system states in this dynamics. The simulations were averaged over 10 realizations of ER networks with $N=10^4$. The white line represents the theory of Eq.\ (5.41) in SI with Eq.\ \eqref{eq:maximum}.  
		(h) The $(\beta,\rho)$ phase diagram for fixed $\Delta=1$. Color represents simulations on ER with $\kappa=20,60,100$ and $N=10^4$. The results were averaged over 10 realizations. The black line stands for the theory of Eq.\ (5.41) in SI with Eq.\ \eqref{eq:maximum}. 
	}
	\label{fig:BrainSpin}		
\end{figure}


\clearpage

\section*{Discussion}

In this paper, we discuss how \emph{dynamics} that take place on a complex network is affected by \emph{dynamic} interventions.
Although a very broad knowledge has been accumulated on the structure of complex networks, the knowledge on dynamics evolving on complex networks is still being discovered. Our study deals with the goal of understanding the ways of \emph{influencing} networks dynamics. 
\vspace{2mm}

We investigate the effect of a simple control of the system, \textit{i.e.}\ forcing a fraction of nodes, $\rho$, to have a desired activity, $x_i=\Delta$. We show that such a simple intervention, even for small $\rho$ and not large $\Delta$, could have a crucial impact on the system dynamic behavior. The control of a fraction of nodes, does not just pull up or down somewhat the natural states of the system, but under certain conditions \emph{eliminates} some of the potential system states.
We show that this elimination of states can transform a functional but \emph{unsustainable} system being \emph{sustainable}. This is because eliminating a potential dysfunctional state by control removes the danger of a transition into a potential undesired inactive state. 
We developed a general framework, applied on three kinds of dynamics, (i) cellular, (ii) neuronal and (iii) spins dynamics, revealing the new phase diagram of controlled system states, by which we can predict, for instance, the minimal fraction of nodes required to make the system sustainable.
\vspace{2mm}

Differently from \emph{control theory} of complex networks \cite{liu2016control}, which explores the ability to move a system into any desired state within a certain continuous volume, here we do not aim to move the system at all, but to eliminate a potential undesired inactive state of the system, and by this, the system stays in its \emph{natural stable active state}, an easier mission allowing our analytical analysis. However, our framework is able to capture a global change in the system's phase diagram, while control theory of nonlinear dynamics on complex networks usually provides only a local information \cite{liu2016control} rather than global.
\vspace{2mm}

Our fundamental and primary analysis for the impact of a simple intervention on network dynamics opens the door to future  studies. 
For instance, 
one could explore other and more complex and/or realistic interventions, as non-random spread of controlled nodes, e.g.\ localized selection, or such as a different control, e.g.\ supplying some flux, constant or dynamic, into the controlled nodes instead of just forcing their activities to be constant as we considered here.
\vspace{2mm}

Note, however, that networks with very broad degree distribution, as scale-free networks with exponent lower than 3, or networks with very low degree, challenge our theory, see Fig.\ \ref{fig:Cellular}a for low $\kappa$, and Fig.\ S4 in SI for SF with $\gamma=2.5$. These challenges demand further research. 

\vspace{2mm}

Finally, it is interesting to note that controlling a very small fraction of nodes can sustain a very large network. This is somewhat analogous to the static structural problem of percolation of interdependent network where a small fraction of reinforced nodes can significantly increase the robustness of the system \cite{yuan2017eradicating}.


\section*{Acknowledgments}

H.S. acknowledges the support of the Presidential Fellowship of Bar-Ilan University,
Israel, and the Mordecai and Monique Katz Graduate Fellowship Program. 
We thank the Israel Science Foundation, the Binational Israel-China Science Foundation (Grant No. 3132/19), the NSF-BSF (Grant No. 2019740), the EU H2020 project RISE (Project No. 821115), the EU H2020 DIT4TRAM, and DTRA (Grant No. HDTRA-1-19-1-0016) for financial support.

\bibliographystyle{unsrt}
\bibliography{bibliography}

\end{document}


\title{ \color{blue} \huge  \vspace{40mm} \bf Sustaining a network by controlling \\[7pt] a fraction of nodes \\ \vspace{10mm} \LARGE Supplementary information}

\maketitle
\thispagestyle{empty}
\clearpage

\tableofcontents
\thispagestyle{empty}
\clearpage
\pagenumbering{arabic}

\clearpage

\section{Free system states}
\label{SecSteadyStateAnalysis}

We consider a class of systems captured by Barzel-Barab\'asi \cite{Barzel2013a,Harush2017,Hens2019}

\begin{equation}
\dod{x_i}{t} = M_0(x_i) + \lambda \sum_{j = 1}^N \m Aij M_1(x_i) M_2(x_j),
\label{Dynamics}
\end{equation}

where $x_i(t)$ is node $i$'s dynamic \textit{activity} ($i = 1,\dots,N$) and the nonlinear functions $M_0(x)$, $M_1(x)$, $M_2(x)$ describe the system's intrinsic dynamics, \textit{i.e}.\ its self-dynamics ($M_0$) and its interaction mechanisms ($M_1, M_2$). The patterns of connectivity between the nodes are captured by the network $\m Aij$, a binary $N \times N$ adjacency matrix, which we assume to follow the configuration model framework, namely a random network with an arbitrary degree distribution, $p_k$. The strength of interactions is governed by the uniform positive parameter $\lambda$. We focus on constructive interactions, in which nodes positively impact each other's activity, or attractive interactions, in which a node attracts its neighbors towards its state. We do not consider competitive interactions.   

The steady-states $\x_{\alpha} = (x_{\alpha,1},\dots,x_{\alpha,N})^{\top}$ of Eq.\ (\ref{Dynamics}) are obtained by setting its derivative on the left hand side to zero and satisfying linear stability. We focus here on systems with at least two steady-states, \textit{i.e}.\ $\alpha = 0,1$, one of which is desirable ($\alpha = 1$) and the other - undesirable ($\alpha = 0$). These states provide an $N$-dimensional descriptions of the system, capturing the stable activity $x_{\alpha,i}$ of all nodes $i = 1,\dots,N$.

To obtain a macroscopic map of the different dynamic states of the system we use the degree-based mean-field approach described in \cite{PastorSatorras2001prl,boguna2002epidemic,Barrat2008,Dorogovtsev2008,PastorSatorras2015}.
Assuming nodes with same degrees behave alike, according to this mean-field approximation, we replace the binary value of $\m Aij$ by its mean value which is the probability of being a connection between $i$ and $j$ given their degrees. In the configuration-model networks which we discuss, this probability equals to $k_ik_j/(N\av{k})$. Thus,
\begin{equation} \label{DBMF1}
	\dod{x_i}{t} = M_0(x_i) + \lambda M_1(x_i) \sum_{j = 1}^N \frac{k_i k_j}{N\av{k}}  M_2(x_j),
\end{equation}
In a relaxation, it turns to
\begin{equation}
	0 = M_0(x_i^*) + \lambda M_1(x_i^*) k_i\sum_{j = 1}^N \frac{ k_j}{N\av{k}}  M_2(x_j^*),
	\label{Relaxation1}
\end{equation}
where $x_i^*$ is $x_i(t\to\infty)$, the activity of node $i$ at steady state.
Here, we define the order parameter $\Theta$ as the sum on the right hand side of \eqref{Relaxation1}, which represents the average impact that a node gets from its neighbor,
\begin{equation}
	\Theta = \sum_{j = 1}^N \frac{ k_j}{N\av{k}}  M_2(x_j^*).
	\label{ThetaFree}
\end{equation}
Hence, by substituting \eqref{ThetaFree} in \eqref{Relaxation1}, we obtain
\begin{equation}
	0 = M_0(x_i^*) + \lambda k_i M_1(x_i^*) \Theta ,
\end{equation}
or
\begin{equation}
	R(x_i^*) = \lambda k_i \Theta ,
	\label{RxiThetaFree}
\end{equation}
where 
\begin{equation}
	R(x)=-\frac{M_0(x)}{M_1(x)} .
	\label{R}
\end{equation}
For invertible $R$, we substitute \eqref{RxiThetaFree} in \eqref{ThetaFree}, to get a self-consistent equation for the order parameter $\Theta$,
\begin{equation}
	\Theta = \sum_{j=1}^{N} \frac{k_j}{N\av{k}}  M_2(R^{-1}(\lambda k_j\Theta)) .
\end{equation}
This equation is solvable for a given formed network which has $k_i$ for $i=1,2,...,N$, but it is not solvable theoretically for a given degree distribution $p_k$.
However, since the term in the summation depends only on $k$ we move to averaging over $k$,
\begin{equation}
	\Theta = \sum_{k} \frac{kp_k}{\av{k}}  M_2(R^{-1}(\lambda k\Theta)) .
	\label{ThetaSCFree}
\end{equation}
We can solve this equation numerically for any degree distribution, $p_k$, and get the solutions for $\Theta$. The solutions of $\Theta$ depend on the connectivity of the network represented by the interactions strength, $\lambda$, and the full degree distribution, $p_k$. 

Next,
to get more simple and universal equation we use another approximation. When $k$ is not widely distributed, and/or the functions $M_2$ and $R$ are close to be linear or close to be constants, we insert the average into the functions \cite{Gao2016}, \textit{i.e.}\ $\overline{M_2(\x)}  = M_2(\bar{\x})$ and $\overline{R(\x)} = R(\bar{\x})$. Doing this in Eq.\ \eqref{ThetaFree}, we obtain
\begin{equation} \label{InsAve1}
	\Theta = M_2\bigg(\sum_{j = 1}^N \frac{ k_j}{N\av{k}}  x_j^*\bigg).
\end{equation}
Thus, we define the average activity as
\begin{equation}
	\bar{\x} = \sum_{i=1}^{N} \frac{k_i}{N\av{k}} x_i^*,
	\label{XFree}
\end{equation}
to get
\begin{equation}
	\Theta = M_2(\bar{\x}).
\end{equation}
By applying the average of $\bar{\x}$ on Eq.\ \eqref{RxiThetaFree}, we obtain
\begin{equation}
	 \sum_{i=1}^{N} \frac{k_i}{N\av{k}}R(x_i^*) = \lambda  \sum_{i=1}^{N} \frac{k_i}{N\av{k}}k_i \Theta = \lambda  \frac{\kkk}{\kk} \Theta .
\end{equation}
Using again the approximation of inserting the average into the function $R$, we get
\begin{equation}
	R(\bar{\x}) = \beta M_2(\bar{\x}),
	\label{RXBetaFree}
\end{equation}
where
\begin{equation}
	\beta = \lambda \kappa 
	\label{Beta}
\end{equation}
is the parameter of the connectivity comprising of $\lambda$, the interaction strength, and
\begin{equation}
	\kappa=\frac{\kkk}{\kk},
	\label{Kappa}
\end{equation}
the average degree of a neighbor.
If $M_2(\bar{\x})$ is not zero, then we obtain
\begin{equation}
	\beta =\frac{ R(\bar{\x})}{M_2(\bar{\x})} ,
	\label{BetaXFree}
\end{equation} 
a simple relation between the average activity in the fixed points and the parameter $\beta$ representing the network connectivity.


\begin{Frame}
\noindent
\textbf{Box I.\ Example.\ Cellular dynamics}.\
As an example we consider gene-regulatory dynamics, captured by the Michaelis-Menten model \cite{Alon2006}, for which (\ref{Dynamics}) takes the form

\begin{equation}
\dod{x_i}{t} = -Bx_i^a + \lambda \sum_{j = 1}^N \m Aij \dfrac{x_j^h}{1 + x_j^h}.
\label{MM}
\end{equation} 

This dynamical equation maps into (\ref{Dynamics}) through $M_0(x) = -Bx^a$, $M_1(x) = 1$ and $M_2(x) = x^h/(1 + x^h)$, thus, $R(x)=Bx^a$, Eq.\ \eqref{R}.
Using the mean-field approximation of (\ref{RXBetaFree}), we write
\begin{equation}
B \bar{\x}^a =  \beta \frac{\bar{\x}^{h}}{1 + \bar{\x}^{h}}.
\label{MMFixedPoint}
\end{equation}
Setting $B = 1,\ a = 1$ and $h = 2$ (see Ref.\ \cite{Gao2016} for a more general treatment), we extract $\bar{\x}$ from \eqref{MMFixedPoint}, predicting three fixed-points as shown in Fig.\ \ref{fig:Cellular}a:\ The inactive (undesirable, red)
\begin{equation}
	\bar{\x}_0 = 0 ,
\end{equation} 
the active (desirable, green)
\begin{equation}
	\bar{\x}_1 = \beta/2 + \sqrt{(\beta/2)^2 - 1} ,
\end{equation} 
and the intermediate (gray dashed-line)
\begin{equation}
	\bar{\x}_2 = \beta/2 - \sqrt{(\beta/2)^2 - 1} .
\end{equation}
From this, we can directly obtain the system's state phase-diagram. The inactive state, $\bar{\x}_0 = 0$, is unconditionally stable. The active state $\bar{\x}_1$ is stable and exists for $\beta\geq2$. The mid-solution $\bar{\x}_2$ is unstable, and hence it does not represent a potential state of the system (Fig.\ \ref{fig:Cellular}a). Therefore $\beta < 2$ forces a collapse on to $\x_0$, while $\beta \geq 2$ represents a bi-stable phase, where the system can reside in both $\x_0$ and $\x_1$, depending on initial conditions. As shown in Fig.\ \ref{fig:Cellular} the transition is located where the curve of $\beta(\bar{\x})$ is twisted. Hence, for finding the critical point of the transition between states, we just have to demand on Eq.\ \eqref{MMFixedPoint},
\begin{equation} \label{MaximumFree}
	\frac{\partial \beta}{\partial \bar{\x}}\bigg|_{\beta_c} = 0.
\end{equation}
providing $\beta_c=2$ and $\bar{\x}_c=1$ (for $B=1,\ a=1,\ h=2$).
\end{Frame}


In \textbf{Box I} we demonstrate this analysis on cellular dynamics, obtaining the potential states and transitions of the gene-regulation, a specific system within our general Eq.\ (\ref{Dynamics}). Cellular dynamics, we find, shows a stable active state only if the adjacency matrix $A$ is sufficiently dense (large $\kappa$) or if the interaction strength is sufficiently strong (large $\lambda$), together satisfying 
\begin{equation}
\beta \geq 2.
\end{equation}
More broadly, this example illustrates the fundamental premise of our present analysis $\bullet$ The potential states of the system and the critical transitions between them, \textit{i.e}.\ the system's phase diagram, are determined by its intrinsic dynamics $M_0(x),M_1(x)$ and $M_2(x)$ $\bullet$ The specific state of a given system, however, \textit{i.e}.\ where it resides along that phase diagram, is predicted by the weighted network topology via $\m Aij$, as encapsulated within $\kappa$ (\ref{Kappa}), and via $\lambda$ $\bullet$ A bi-stable regime implying that even when a system resides in the active state, it is not completely safe since there exists also the potential inactive state, therefore we define such a	 state as unsustainable, as we discuss below.

\subsection{Sustainable and unsustainable states}
\begin{figure}[h]
	\centering
	\includegraphics[width=0.75\linewidth]{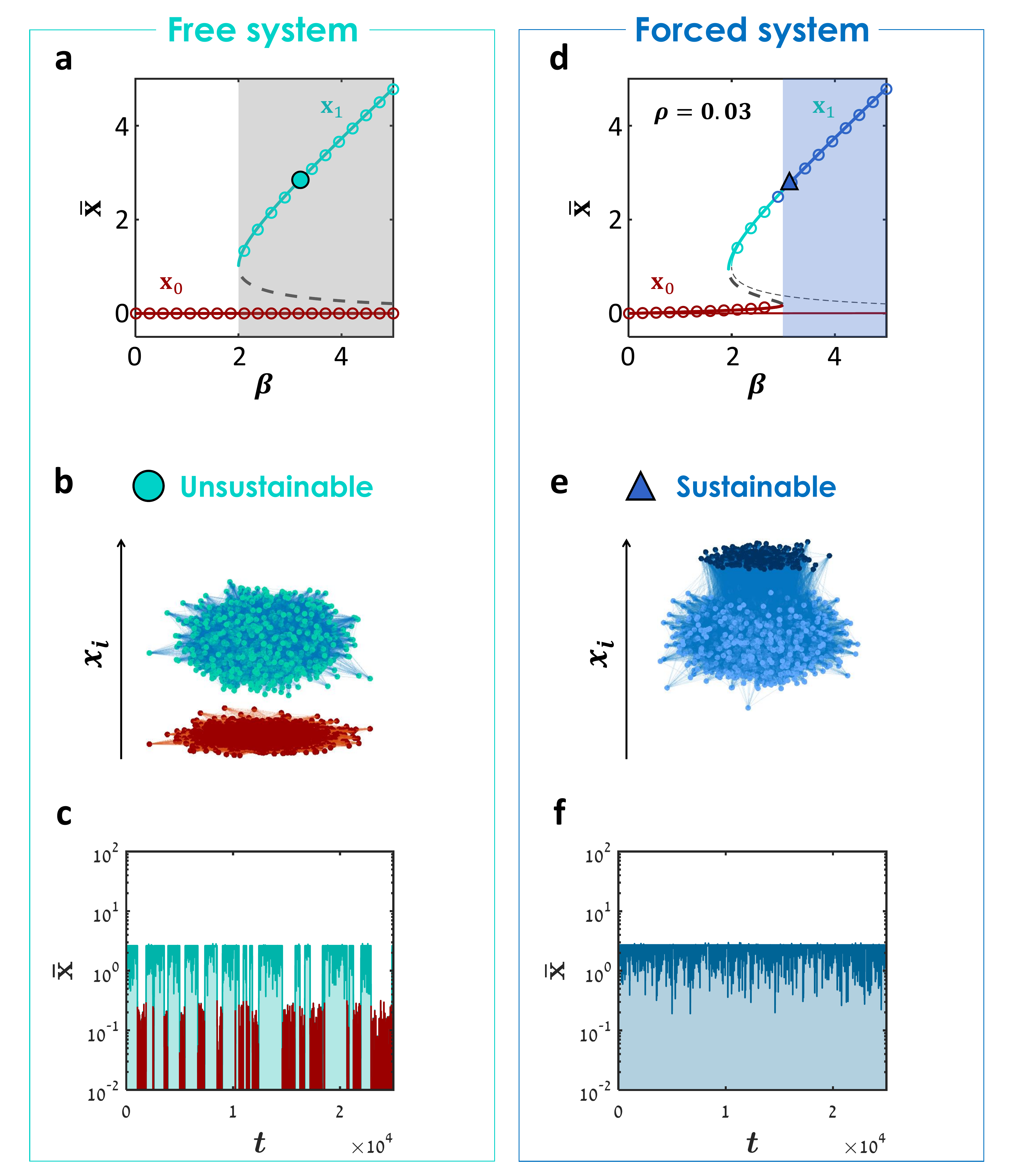}
\caption{	
	\textbf{Sustainability of a cellular network.} 	
	(a) Simulations (symbols) and theory (lines, Eq.\ \eqref{MMFixedPoint} with $\rho=0$) results for a free system with ER structure with $N=10^4$ and $\kappa=40$. The system has a bi-stable region (gray shade).
	(b) The activities of a system with $\beta=3.1$ for $\rho=0$. Both states, active and inactive, are stable, therefore the system is unsustainable.
	(c) Demonstration of the unsustainable character of the system in (b) for $\beta=3.1$, and $\rho=0$. Adding a noise to the activities $x_i$ causes the system to transit randomly between both states, the active (green) and the inactive (red), since both are stable. Thus, this system can be regarded as unsustainable.
	(d) For a forced system by a fraction $\rho=0.03$ of random nodes held with activity $\Delta=5$, Eqs.\ \eqref{BetaXForced} or \eqref{BetaXForcedMM} provide the phase diagram (thick curve), exhibiting an s-shape curve which has now also a regime with only a single active state (blue shade). This regime is the sustainable phase. Note that simulations (symbols) are in good agreement with the theory. The network is the same as in (a).
	(e) Activities for $\beta=3.1$ as in (b) but with control of $\rho=0.03$ and $\Delta=5$. The dark blue nodes are the forced nodes, and the light blue nodes represent the sustainable $\x_1$, since the inactive state $\x_0$ disappears due to the intervention.
	(f) Demonstration of the sustainable system in (e). We added the same noise as in (c), but here the system does not collapses into a dysfunctional state since there is no such a state due to the control.
} 
\label{fig:Cellular}	
\end{figure}

The structure of the phase diagram yielded by Eq.\ \eqref{BetaXFree} (Fig.\ \ref{fig:Cellular}a) indicates that for cellular dynamics the inactive state $\x_0$ is steady regardless of the value of $\beta$. Hence, for large $\beta\geq2$ where the active state $\x_1$ exists, even if the system resides at $\x_1$, it is \emph{unsustainable}, since the inactive state is steady as well (see demonstrations in Fig.\ \ref{fig:Cellular}b,c), and thus the system is not in a safe state and might move to $\x_0$ under some perturbations in its activity. We define as $sustainable$ a system residing at an active state for which a potential inactive state does not exist for the given network structure ($A$) and interaction strength ($\lambda$). System sustainability, therefore, mandates dynamic intervention or \textit{sustaining}, as we derive in Sec.\ \ref{SecSustainability}. A bi-stable regime where systems have both states for the same network, arise quite commonly in dynamics within the form (\ref{Dynamics}). We therefore, seek to characterize our ability to dynamically sustain an unsustainable system via parsimonious dynamic interventions.

\clearpage

\section{Sustaining a network}
\label{SecSustainability}

\subsection{Modeling system sustaining}

Consider a system of the type discussed above, characterized by two stable states - an undesirable $\x_0$ and a desirable $\x_1$, see Fig.\ \ref{fig:Cellular}. Let us further assume that the system is in the bi-stable phase, presently at the desirable $\x_1$. We seek dynamic interventions that will help us prevent the system from potentially moving to the undesirable state $\x_0$ by eliminating it. To achieve this, we assign a selected set of nodes $\mathcal{F}$ - the \textit{forced} nodes - whose dynamics we externally control. The complementary set of $\F$, containing the free dynamic nodes is denoted by $\D=\F^C$. For simplicity, but without loss of generality, we consider the simplest control, that is forcing the activity of nodes in $\F$ to be in some uniform constant high value $\Delta$. This control, effectively, changes the system's dynamics, Eq.\ \eqref{Dynamics}, into
\begin{equation}
\left\{
\begin{array}{cclr}
x_i(t) &=& \Delta & i \in \mathcal{F}
\\[10pt]
\dod{x_i}{t} &=& M_0(x_i) + \lambda \displaystyle \sum_{j = 1}^N \m Aij M_1(x_i) M_2(x_j) & i \in \D
\end{array}
\right.,
\label{ForcedDynamics}
\end{equation}
in which nodes in $\mathcal{F}$ are forced to follow the external control, while the remaining $|\D|=N-|\mathcal{F}|$ nodes continue to evolve via the system's natural interaction dynamics. The fraction of controlled nodes is denoted as \begin{equation}
	\rho = \frac{|\F|}{N}.
\end{equation}
To follow the impact of such a control, and testing if it makes the system sustainable, we analyze Eq.\ \eqref{ForcedDynamics} for finding system's states while it is forced by the intervention described above.


\subsection{Forced system analysis}
\label{SecForced}

Let us observe the dynamics of the free nodes ($i\in\D$) in Eq.\ \eqref{ForcedDynamics},
\begin{equation}
	\dod{x_i}{t} = M_0(x_i) + \lambda \sum_{j = 1}^N \m Aij M_1(x_i) M_2(x_j),
\end{equation}
which by separating the sum between neighbors in $\D$ and neighbors in $\F$ turns to
\begin{equation}
	\dod{x_i}{t} = M_0(x_i) + \lambda M_1(x_i) \bigg(\sum_{j \in \D} A_{ij}  M_2(x_j) + \sum_{j \in \F} A_{ij}  M_2(\Delta) \bigg).
\end{equation}
To find the steady states of the system, we demand a relaxation, thus the derivative vanishes,
\begin{equation}
	0 = M_0(x_i^*) + \lambda M_1(x_i^*) \bigg(\sum_{j \in \D} A_{ij}  M_2(x_j^*) + k_{i}^{\D\to\F}  M_2(\Delta) \bigg),
\end{equation}
where $x_i^*$ represents the relaxation activity value of node $i$, and $k_i^{\D\to\F}$ denotes the number of neighbors in $\F$ of the node $i$ which is in $\D$.
Arranging the terms, we obtain
\begin{equation}
	R(x_i^*) = \lambda \bigg(\sum_{j \in \D} A_{ij}  M_2(x_j^*) + k_{i}^{\D\to\F}  M_2(\Delta) \bigg),
\end{equation}
where $R(x)=-M_0(x)/M_1(x)$ as defined in Eq.\ \eqref{R}. 
Next, we apply the Degree-Based Mean-Field approximation \cite{PastorSatorras2001prl,boguna2002epidemic,Barrat2008,Dorogovtsev2008,PastorSatorras2015}, replacing the binary value, $\m Aij$, by the probability of $i\in\D$ and $j\in\D$ to be connected given their degrees, which is $k_i^{\D\to\D}k_j^{\D\to\D}/(|\D|\av{k_{\D\to\D}})$ because of the configuration model structure. This gives
\begin{equation}
	R(x_i^*) = \lambda \bigg(\sum_{j \in \D} \frac{k_i^{\D\to\D}k_j^{\D\to\D}}{|\D|\av{k_{\D\to\D}}}  M_2(x_j^*) + k_{i}^{\D\to\F}  M_2(\Delta) \bigg).
\end{equation}
Note that now the sum does not depend on $i$, hence we can write
\begin{equation}
	R(x_i^*) = \lambda k_i^{\D\to\D} \Theta  + \lambda k_{i}^{\D\to\F}  M_2(\Delta),
	\label{RxThetaForced}
\end{equation}
by defining the order parameter,
\begin{equation}
	\Theta =  \frac{1}{|\D|\av{k_{\D\to\D}}} \sum_{j \in \D} k_j^{\D\to\D}  M_2(x_j^*) ,
	\label{ThetaForced}
\end{equation}
which is the average impact of a node in $\D$ (free node) on a node in $\D$. Note that this definition for $\Theta$ includes the above definition, Eq.\ \eqref{ThetaFree}, for a free system since $\D$ is all the network in a free system. 
Using Eqs.\ \eqref{RxThetaForced} and \eqref{ThetaForced}, we obtain a single self-consistent equation for the order parameter $\Theta$,
\begin{equation}
	\Theta =  \frac{1}{|\D|\av{k_{\D\to\D}}}  \sum_{j \in \D} k_j^{\D\to\D}  M_2(R^{-1}(\lambda k_j^{\D\to\D} \Theta  + \lambda k_{j}^{\D\to\F}  M_2(\Delta))) .
	\label{ThetaSC1}
\end{equation}
To solve this equation, we should replace the summation on nodes in $\D$ by some theoretically calculable term without need to measure nodes degree. Thus, we look for different expression where Eq.\ \eqref{ThetaSC1} becomes
\begin{equation}
	\Theta =  \frac{1}{\av{k_{\D\to\D}}} \left\langle k^{\D\to\D}  M_2(R^{-1}(\lambda k ^{\D\to\D} \Theta  + \lambda k^{\D\to\F}  M_2(\Delta))) \right\rangle.
\end{equation}
Averaging over all the possibilities of the pair $k^{\D\to\D}$ and $k^{\D\to\F}$, it gets the form
\begin{equation}
	\Theta = \frac{1}{\av{k_{\D\to\D}}}\sum_{k,k'} k \Pr(k_j^{\D\to\D}=k,k_j^{\D\to\F}=k') M_2(R^{-1}(\lambda k \Theta  + \lambda k'  M_2(\Delta))),
\end{equation}
where $\Pr(k_j^{\D\to\D}=k,k_j^{\D\to\F}=k')$ is the joint probability that node $j\in\D$ has $k$ neighbors in $\D$ (free nodes) and $k'$ neighbors in $\F$ (forced nodes). This joint probability is determined by the degree distribution $p_k$ and the way of selection of the forced nodes ($\F$). Next, instead of running over $(k,k')$ (which are $k^{\D\to\D}$ and $k^{\D\to\F}$ respectively), we run over $(k_0,k)$ where here $k_0=k^{\D\to\V}$ is the degree of a free node, where $\V$ is the set of all nodes, and $k=k^{\D\to\D}$ is the number of the free neighbors of a free node. Of course for each node $k'=k_0-k$, that is $k_i^{\D\to\F}=k_i^{\D\to\V}-k_i^{\D\to\D}$.
Performing this change of indexes, we obtain
\begin{equation}
	\Theta = \frac{1}{\av{k_{\D\to\D}}} \sum_{k_0} \sum_{k \leq k_0} k \Pr(k,k_0) M_2(R^{-1}(\lambda k \Theta  + \lambda (k_0-k)  M_2(\Delta))),
\end{equation}
and substituting $\Pr(k,k_0)=\Pr(k_0)\Pr(k|k_0)$ yields
\begin{equation}
	\Theta = \frac{1}{\av{k_{\D\to\D}}} \sum_{k_0} \Pr(k_0)\sum_{k \leq k_0} k \Pr(k|k_0) M_2(R^{-1}(\lambda k \Theta  + \lambda (k_0-k)  M_2(\Delta))) .
	\label{ThetaSC5}
\end{equation}
The terms $\av{k_{\D\to\D}}$, $\Pr(k_0)$, and $\Pr(k|k_0)$ depend on the way of selection of the forced nodes besides the degree distribution $p_k$. For now, we consider a completely \emph{random selection} of controlled nodes. In this case, $\Pr(k_0)=p_{k_0}$ (the degree distribution of the network) because $\D$ is random and therefore has the same degree distribution $p_k$ as the whole network $\V$. The probability that a neighbor is in $\F$ is $\rho$, thus $\av{k_{\D\to\D}}=\av{k_{\V\to\D}}=\kk(1-\rho)$ as an average of binomial distribution, and $\Pr(k|k_0)=\binom{k_0}{k}(1-\rho)^k \rho^{k_0-k}$, binomial distributed. Substituting these, we finally obtain for a random selection of controlled nodes,
\begin{equation}
	\Theta = \frac{1}{\av{k}(1-\rho)} \sum_{k_0} p_{k_0}\sum_{k \leq k_0} k \binom{k_0}{k}(1-\rho)^k \rho^{k_0-k} M_2(R^{-1}(\lambda k \Theta  + \lambda (k_0-k)  M_2(\Delta))) ,
	\label{ThetaSCForcedSol}
\end{equation}
where $p_{k_0}$ is the degree distribution and $\kk$ is the mean degree. This self-consistent equation is solvable theoretically based on the degree distribution. We use Eq.\ \eqref{ThetaSCForcedSol} to plot the theoretical lines in Fig.\ 4 in the main text.


\subsubsection{Small fluctuations MF}

To get a simpler and informative equation, we apply another Mean-Field approximation \cite{Gao2016} that works well for small fluctuations as we discuss below. According to this MF, we insert the average in Eq.\ \eqref{ThetaForced} into the function $M_2(x)$, to obtain
\begin{equation}
	\Theta =  M_2\bigg(\frac{1}{|\D|\av{k_{\D\to\D}}} \sum_{j \in \D} k_j^{\D\to\D}  x_j^*\bigg) = M_2(\bar{\x}),
	\label{ThetaXForced}
\end{equation}
where the average activity of the unforced nodes, $\bar{\x}$, is defined by
\begin{equation}
	\bar{\x} = \frac{1}{|\D|\av{k_{\D\to\D}}} \sum_{j \in \D} k_j^{\D\to\D}  x_j^* .
	\label{XForced}
\end{equation}
This definition includes Eq.\ \eqref{XFree} because for free system $\D=\V$, \textit{i.e.}\ all the network.
Operating this average on Eq.\ \eqref{RxThetaForced} and using the MF approximation, we get
\begin{equation}
	R(\bar{\x}) =  \lambda \kappa_{\D\to\D} \Theta  + \lambda \kappa_{\D\to\F}  M_2(\Delta),
	\label{RXThetaForced}
\end{equation}
where 
\begin{equation} \label{kappaDDkappaDF}
	\begin{aligned}
		\kappa_{\D\to\D} &=  \dfrac{\av{k^2_{\D\to\D}}}{\av{k_{\D\to\D}}},
		\\[7pt]
		\kappa_{\D\to\F} &= \dfrac{\av{k_{\D\to\D}k_{\D\to\F}}}{\av{k_{\D\to\D}}},
	\end{aligned}
\end{equation}
are the average degrees into $\D$ and into $\F$ respectively of node approached by link within $\D$. 
Substituting Eq.\ \eqref{ThetaXForced} into Eq.\ \eqref{RXThetaForced} we obtain an equation for the activity of a forced system,
\begin{equation}
	R(\bar{\x}) =  \lambda \kappa_{\D\to\D} M_2(\bar{\x}) + \lambda \kappa_{\D\to\F}  M_2(\Delta).
	\label{XForced1}
\end{equation}
The quantities $\kappa_{\D\to\D}$ and $\kappa_{\D\to\F}$ depend of course on the way of selecting the set $\F$ of controlled nodes.
For now, we assume for simplicity that $\F$ is selected completely randomly, and find these quantities for this case. Below in Section \ref{SecArbDegDist} we analyze the general case.

\subsubsection{The approximation $\overline{M_2(\x)}  = M_2(\bar{\x})$ and $\overline{R(\x)} = R(\bar{\x})$ }
\label{SecMbarxRbarx}

The approximation we used above is exact if at least one of the following two conditions applies:\ (i) $M_2(x)$ and $R(x)$ are linear; (ii) $x_i(t)$ are uniform. Clearly, these conditions are not guaranteed, however, under many practical scenarios, they represent a sufficient approximation, designed to detect the macro-scale behavior of the system - as fully corroborated by our numerical examination. Indeed, while Eq.\ (\ref{Dynamics}) is, generally, nonlinear, $M_2(x), R(x)$, in many useful models, are often sub-linear, linear or weakly super-linear, \textit{i.e}.\ involving powers that are not much higher than unity. This satisfies, approximately, condition (i). In other cases we may observe strong nonlinearities in $M_2(x)$ and $R(x)$, but in such cases, we often have bounded activities $x_i(t)$. This ensures a narrow distribution of $x_i(t)$, roughly satisfying condition (ii). We further elaborate on the relevance of these conditions in the appropriate sections, where we analyze each of our specific dynamic systems, Sec. \ref{SecDynamicModels}.

\subsubsection{Random selection} \label{SecRandom}
Selecting the set of controlled nodes $\F$ randomly determines that the degree distribution of the controlled nodes, $\F$, the free nodes $\D$, and all of the nodes $\V$, are the same, $p_k$. This allows us to analyze the quantities in Eq.\ \eqref{kappaDDkappaDF}, appearing in Eq.\ \eqref{XForced1}, since we can replace the average over subset of the network ($\F$ or $\D$) by an average over all the network $\V$. Considering that an arbitrary node in $\V$ belongs to $\F$ with likelihood $\rho$, we obtain, using Wald's identity,
\begin{equation}
	\av{k_{\D\to\D}} = \av{k_{\V\to\D}}=\kk(1-\rho).
	\label{kDD}
\end{equation}
For the second moment $\av{k^2_{\D\to\D}}$, we first change the average to be on $\V$, $\av{k^2_{\V\to\D}}$. Then we define a variable $y_j$ which is an indicator of whether neighbor number $j$ of a given node belongs to $\D$, such that $y_j=1$ if neighbor $j$ belongs to $\D$ and $y_j=0$ otherwise. Thus, $k_{\V\to\D}=\sum_{j=1}^{k}y_j$ where the index $j$ runs over neighbors of an arbitrary node. The random variables $y_j$ are independent, and have the mean $1-\rho$, therefore, 
\begin{equation}
	\begin{aligned}
		\av{k^2_{\D\to\D}} & = \left\langle k^2_{\V\to\D} \right\rangle = \Big\langle \Big( \sum_{j=1}^k y_j \Big)^2 \Big\rangle=\Big\langle \sum_{j=1}^k y_j^2+2\sum_{1\leq i< j \leq k} y_iy_j \Big\rangle
		\\
		& = \kk(1-\rho)+\av{k^2-k}(1-\rho)^2 .
	\end{aligned}
	\label{k2DD}
\end{equation}
Substituting Eqs.\ \eqref{kDD} and \eqref{k2DD} in Eq.\ \eqref{kappaDDkappaDF} provides
\begin{equation}
		\kappa_{\D\to\D} =  \dfrac{\av{k^2_{\D\to\D}}}{\av{k_{\D\to\D}}} = 1+(\kappa-1)(1-\rho).
		\label{kappaDD}	
\end{equation}
This result is very intuitive because this quantity represents the expected number of free neighbors of a free node approached via free node as well. Hence, it has certainly one free neighbor, and since the average degree of an arbitrary neighbor is $\kappa$, it has $\kappa-1$ more neighbors but a fraction $\rho$ of them is expected to be controlled.

Next, we move to evaluate the second term in Eq.\ \eqref{kappaDDkappaDF}.
\begin{equation}
	\begin{aligned}
		\av{k_{\D\to\D}k_{\D\to\F}} &= \av{k_{\V\to\D}k_{\V\to\F}} 
		\\[5pt]
		& = \av{k_{\V\to\D}(k-k_{\V\to\D})}
		\\[5pt]
		& =\av{k_{\V\to\D}k}-\av{k^2_{\V\to\D}} .
	\end{aligned} 
\label{kDDkDFav}
\end{equation}
We already found the second term in Eq.\ \eqref{k2DD}. The first term is
\begin{equation}
	\begin{aligned}
		\av{k_{\V\to\D}k} & = \bigg \langle \Big( \sum_{j=1}^k y_j \Big)\Big( \sum_{i=1}^k 1 \Big) \bigg \rangle
		\\[5pt]
		& = \bigg \langle \sum_{i=1}^k \sum_{j=1}^k y_j \bigg \rangle  
		= \kkk(1-\rho) .
	\end{aligned}
	\label{kVDkAv}
\end{equation}
Plugging Eqs.\ \eqref{k2DD} and \eqref{kVDkAv} into Eq.\ \eqref{kDDkDFav} gives
\begin{equation}
	\begin{aligned}
		\av{k_{\D\to\D}k_{\D\to\F}} 
		& =\av{k_{\V\to\D}k}-\av{k^2_{\V\to\D}} 
		\\[5pt]
		& = \kkk(1-\rho)-(\kk(1-\rho)+\av{k^2-k}(1-\rho)^2 )
		\\[5pt] 
		& = (\kkk-\kk)(1-\rho)\rho.
	\end{aligned}
	\label{kDDkDFAv2}
\end{equation}
Substituting Eqs.\ \eqref{kDD} and \eqref{kDDkDFAv2} into Eq.\ \eqref{kappaDDkappaDF} we get
\begin{equation}
	\kappa_{\D\to\F} =  \dfrac{\av{k_{\D\to\D}k_{\D\to\F}}}{\av{k_{\D\to\D}}} = (\kappa-1)\rho.	
	\label{kappaDF}
\end{equation}
This formula is intuitive as well, since it represents the number of forced neighbors of a free node approached via another free node. $\kappa$ is the expected degree, and among the remain $\kappa-1$ neighbors, a fraction $\rho$ is expected to be forced. \\
Substituting Eqs.\ \eqref{kappaDD} and \eqref{kappaDF} into Eq.\ \eqref{XForced1} we get the Mean-Field equation for a forced system for a random selection of controlled nodes,
\begin{equation}
	R(\bar{\x}) = \lambda (1+(\kappa-1)(1-\rho))   M_2(\bar{\x}) + \lambda (\kappa-1)\rho M_2(\Delta).
	\label{ForcedMFrandom}
\end{equation}
Taking the limit $\kappa \gg 1$ and recalling $\beta=\lambda\kappa$, this equation takes the form
\begin{equation}
	R(\bar{\x}) = \beta(1-\rho) M_2(\bar{\x}) + \beta\rho M_2(\Delta).
	\label{ForcedMFRandomBeta}
\end{equation}
Isolating $\beta$, we finally obtain Eq.\ (8) in the main text,
\begin{equation}
	\beta = \frac{ R(\bar{\x})}
	{(1-\rho)M_2(\bar{\x})+\rho M_2(\Delta)} ,
	\label{BetaXForced}
\end{equation}
which yields a relation between $\bar{\x}$ and $\beta$, determining the average activity of the forced system for each topology captured by $\beta$. Also the system states depend on the intervention characteristics, $\rho$ and $\Delta$. 
Note that substituting $\rho=0$ in Eq.\ \eqref{BetaXForced} recovers Eq.\ \eqref{BetaXFree} of a free system.
Recognizing the stable states of forced system compared to those of free system, we can answer our main question of sustaining a system by holding a fraction of nodes. In the following Section we find a region where unsustainable network becomes sustainable due to the external control. We call this region the sustainability window.

\subsection{Sustainability window}
\label{SecWin}

From Eq.\ \eqref{BetaXForced} we get a new phase diagram for a forced given $\rho$ and $\Delta$, which is different from the free system's phase diagram. When there is a range of $\beta$-values in which the free system is unsustainable, \textit{i.e.}\ there exist both the active and the inactive states ($\x_1$ and $\x_0$), and in the same region of $\beta$-values the forced system is sustainable, \textit{i.e.}\ there exists only $\x_1$, then this range is regarded as a sustainability window. This is because a system residing in this region becomes sustainable by controlling a fraction $\rho$ with force $\Delta$.

In Fig.\ \ref{fig:Cellular} we demonstrate this control effect on gene regulation (elaborately analyzed below in Section \ref{SecDynamicModels}). Fig.\ \ref{fig:Cellular}d shows the phase diagram of a forced system derived from Eq.\ \eqref{BetaXForced}. The thin light lines in Fig.\ \ref{fig:Cellular}d as well as in Fig.\ \ref{fig:Cellular}a represent the states of a free system. One can see that above some value of $\beta=\beta_c$ the forced system shows only the active state, while a free system has both states. Thus, this region, $\beta>\beta_c$, is the sustainability window (blue shade). Fig.\ \ref{fig:Cellular}e demonstrates the elimination of the inactive state $\x_0$ due to the external control. Fig.\ \ref{fig:Cellular}f demonstrates the implication of the system's being sustainable. When adding large noise to the activities $x_i$, the system never collapses but always goes up into the stable active state. This is in contrary to a free system with same $\beta$, Fig.\ \ref{fig:Cellular}c, which makes transitions between both states due to the random noise.

As seen in Fig.\ \ref{fig:Cellular}d, the value of $\beta_c$, above which the network is sustainable, can be obtained by finding the local maximum of $\beta(\bar{\x})$. Therefore it is found using Eq.\ \eqref{BetaXForced} by
\begin{equation} \label{Maximum}
	\frac{\partial \beta}{\partial \bar{\x}}\bigg|_{\beta_c} = 0.
\end{equation}
Solving Eq.\ \eqref{Maximum}, we find the transition point $(\beta_c,\bar{\x}_c)$.
From Eqs.\ \eqref{BetaXForced} and \eqref{Maximum} it can be seen that $\beta_c$ depends on the fraction $\rho$, and the force of control, $\Delta$. 
Having the relation $\beta_c(\rho,\Delta)$, we derive from it the inverse relation $\rho_c(\beta,\Delta)$, \textit{i.e.}\ the minimal fraction of nodes required to be controlled, given the connectivity $\beta$ and the force $\Delta$, for sustaining the network.

Increasing $\rho$ more unveils another critical value of $\rho=\rho_0$ as explained below, the tricritical point.

\subsection{Tricritical point}

As shown in Figs.\ 3d,e in the main text, increasing $\rho$, shortens the unsustainable interval (the bi-stable regime). At some critical point $\rho=\rho_0$, the unsustainable phase completely vanishes and the transition between the active state, $\x_1$, and the inactive state, $\x_0$, becomes continuous. One can see that at this point the maximum and the minimum of $\beta(\bar{\x})$ merge. This consideration yields the additional condition besides Eq.\ \eqref{Maximum} for the tricritical point,
\begin{equation} \label{Maximum2}
	\frac{\partial^2 \beta}{\partial \bar{\x}^2}\bigg|_{\beta_c} = 0.
\end{equation}
These two conditions, Eqs.\ \eqref{Maximum} and \eqref{Maximum2}, together determine a tricritical point $(\beta_0,\rho_0)$ in $(\beta,\rho)$-space where the three phases (sustainable, unsustainable and inactive) meet, and beyond which the system does not experience an abrupt transition at all. See Section \ref{SecDynamicModels} for explicit calculations for particular dynamics.


\clearpage

\section{Methods of selecting the controlled nodes}
\label{SecArbDegDist}

In this section we analyze several different methods of selecting nodes to sustain the system via controlling them, besides the simple case of random selection that we analyzed above in Sec.\ \ref{SecRandom}.
All selection methods we shall consider, obey the character of random spread throughout the network. However, they differ from each other in the degree distribution of the controlled node. Therefore, we explore the general case of degree distribution of the set $\F$, and then analyze particular cases.

\subsection{Arbitrary degree distribution of controlled nodes}

Let the degree distribution of the forced nodes be $f_k$, while the degree distribution of the whole network is $p_k$.
The forced nodes are distributed randomly over the network, excluding \textit{e.g.}\ a localized selection.
We aim to modify the above equations predicting the state a forced system, which were obtained for a random selection, for a general degree distribution. Hence, we wish to update both Eq.\ \eqref{ThetaSCForcedSol}, used for large fluctuations, and Eq.\ \eqref{BetaXForced}, used for small fluctuations. 

To this end, we have to step back to the stages before we assumed that the selection is random. For large fluctuations, we return to Eq.\ \eqref{ThetaSC5} and we recalculate the quantities $\av{k_{\D\to\D}}$, $\Pr(k_0)$ and $\Pr(k|k_0)$. For small fluctuations, we go back to Eq.\ \eqref{XForced1}, and we recalculate $\kappa_{\D\to\D}$ and $\kappa_{\D\to\F}$ defined in Eq.\ \eqref{kappaDDkappaDF}.
We will follow step by step until we obtain all these quantities, then we rewrite the equations for a forced system for a selection of $\F$ with arbitrary degree distribution $f_k$.

First, we find the probability that a random node approached via a free neighbor is forced. We assume that there are no correlations between the degrees according to the configuration model.
Using Bayes' theorem for the likelihood of an arbitrary node to be forced and to have degree $k$, we get
\begin{equation}
	\Pr(k)\Pr(\F|k) = \Pr(\F)\Pr(k|\F),
\end{equation}
where $\Pr(\F|k)$ is the probability that an arbitrary node belongs to $\F$ given its degree is $k$, and $\Pr(k|\F)$ is the opposite.
Three of the terms in the last equation are already known,
\begin{equation}
	p_k\Pr(\F|k) = \rho f_k.
\end{equation}
Thus, we obtain the probability of a random node with degree $k$ to belong to $\F$, \textit{i.e.} to be forced,
\begin{equation}
	\Pr(\F|k) = \rho f_k / p_k .
\end{equation}
Now, when we approach a node via a neighbor, using the degree distribution of a neighbor $kp_k/\kk$, we know that the chance it is forced, is obtained by
\begin{equation}
	\Pr(\F|{\rm neighbor}) = \sum_{k=0}^{\infty} \frac{kp_k}{\kk}\Pr(\F|k) = \sum_{k=0}^{\infty} \frac{kp_k}{\kk}\rho f_k/p_k = \rho \kk_{\F}/\kk .
\end{equation}
This quantity is useful since the nodes that appear in the interaction terms in Eq.\ \eqref{Dynamics} are approached via their neighbors, hence we denote this probability, $\Pr(\F|{\rm neighbor})$, by $\rho^*$, and get
\begin{equation} \label{RhoStar}
	\rho^*  = \rho \kk_{\F}/\kk .
\end{equation}
For random selection $\rho^*=\rho$, since $\kk_{\F}=\kk$. This is reasonable because there is no correlation between the degree and the identity of the node, \textit{i.e.}\ whether the node is free or forced. However, in degree-dependent selection, since a neighbor has on average larger degree, it is more likely to be forced compared to random node if $\kk_{\F}>\kk$ and vice versa. 

Next, we move to calculate the degree distribution of the subset $\D$ of free nodes. Let us observe the number of nodes with degree $k$ in each set from our three sets (free, forced and all) that fulfill the disjoint union $\D\sqcup\F=\V$,
\begin{equation}
	 |\F|\Pr(k|\F) + |\D|\Pr(k|\D) = N\Pr(k).
\end{equation}
Plugging into this equation what we already know, we get
\begin{equation}
	N\rho f_k + N(1-\rho)\Pr(k|\D) = Np_k,
\end{equation}
resulting in
\begin{equation}
	 \Pr(k|\D) = (p_k - \rho f_k) / (1-\rho).
\end{equation}
This is the distribution that we denoted above as $\Pr(k_0)$.
Knowing this distribution, we can easily calculate its moments,
\begin{equation}
	\begin{array}{lll}
		\kk_{\D} &= & \left(\kk - \rho \kk_{\F} \right) / (1-\rho),	
		\\[10pt]	
		\kkk_{\D} &= & \left(\kkk - \rho \kkk_{\F} \right) / (1-\rho).
	\end{array}	
\end{equation}
Hence, we obtain the expected degree of a neighbor in $\D$,
\begin{equation} \label{kappaD}
	\kappa_{\D} = \dfrac{\kkk_{\D}}{\kk_{\D}} = \frac{\kkk - \rho \kkk_{\F}}{\kk - \rho \kk_{\F}}.
\end{equation}
%
%
Next, we progress to calculate the components of $\kappa_{\D\to\D}$ and $\kappa_{\D\to\F}$ in Eq.\ \eqref{kappaDDkappaDF}. As above for a random selection, Eqs.\ \eqref{kDD}, \eqref{k2DD} and \eqref{kDDkDFav}, we define the variable $y_j$ which equals 1 if neighbor number $j$ of a given node belongs to $\D$ and 0 otherwise. Thus, according to Wald's identity,
\begin{equation}
	\av{k_{\D\to\D}} = \bigg\langle \sum_{j=1}^{k_{\D\to\V}}y_j \bigg\rangle = \av{k_{\D\to\V}}(1-\rho^*) = \kk_{\D}(1-\rho^*).
	\label{kDDArb}
\end{equation}
The quantity $\rho^*$ appears instead of $\rho$ in the random case, since the probability that a neighbor is forced is $\rho^*$ as obtained above, Eq.\ \eqref{RhoStar}.
For the second moment $\av{k^2_{\D\to\D}}$,  
\begin{equation}
	\begin{aligned}
		\av{k^2_{\D\to\D}} & = \Big\langle \Big( \sum_{j=1}^{k_{\D\to\V}} y_j \Big)^2 \Big\rangle=\Big\langle \sum_{j=1}^{k_{\D\to\V}} y_j^2+2\sum_{1\leq i< j \leq k_{\D\to\V}} y_iy_j \Big\rangle
		\\[5pt]
		& = \av{k_{\D\to\V}}(1-\rho^*)+\av{k_{\D\to\V}^2-k_{\D\to\V}}(1-\rho^*)^2 
		\\[15pt]
		& = \kk_{\D}(1-\rho^*)+\av{k^2-k}_{\D}(1-\rho^*)^2 .
	\end{aligned}
	\label{k2DDArb}
\end{equation}

For $\av{k_{\D\to\D}k_{\D\to\F}}$, we obtain,
\begin{equation}
	\begin{aligned}
		\av{k_{\D\to\D}k_{\D\to\F}} &=  \av{k_{\D\to\D}(k_{\D\to\V}-k_{\D\to\D})}
		\\[15pt]
		& =\av{k_{\D\to\D}k_{\D\to\V}}-\av{k^2_{\D\to\D}}
		\\[5pt]
		& = \bigg \langle \Big( \sum_{j=1}^{k_{\D\to\V}} y_j \Big)\Big( \sum_{i=1}^{k_{\D\to\V}} 1 \Big) \bigg \rangle - \av{k^2_{\D\to\D}} 
		\\[0pt]
		& = \bigg \langle \sum_{i=1}^{k_{\D\to\V}} \sum_{j=1}^{k_{\D\to\V}} y_j \bigg \rangle  - \av{k^2_{\D\to\D}}
		\\[5pt]
		& = \av{k_{\D\to\V}^2}(1-\rho^*) - \av{k^2_{\D\to\D}}.
	\end{aligned} 
	\label{kDDkDFArb}
\end{equation}
Substituting Eq.\ \eqref{k2DDArb} into Eq.\ \eqref{kDDkDFArb} yields
\begin{equation}
	\begin{aligned}
		\av{k_{\D\to\D}k_{\D\to\F}} 		
		& = \kkk_{\D}(1-\rho^*) - \left(\kk_{\D}(1-\rho^*)+\av{k^2-k}_{\D}(1-\rho^*)^2 \right)
		\\[5pt] 
		& = \left(\kkk_{\D}-\kk_{\D}\right)(1-\rho^*)\rho^*.
	\end{aligned}
	\label{kDDkDF2Arb}
\end{equation}

Using Eqs.\ \eqref{kDDArb}, \eqref{k2DDArb} and \eqref{kDDkDF2Arb}, we get the version of Eq.\  \eqref{kappaDDkappaDF} for an arbitrary degree distribution of $\F$,
\begin{equation}
	\begin{aligned}
		\kappa_{\D\to\D} &= 1+(\kappa_{\D}-1)(1-\rho^*),
		\\[10pt]
		\kappa_{\D\to\F} &= \left(\kappa_{\D}-1\right)\rho^*.
	\end{aligned}	
	\label{kappaDDkappaDFarbitrary}	
\end{equation}
This result is again intuitive since our equation deals with an average free node which was approached by neighbor in $\D$. Therefore its expected degree is $\kappa_{\D}$. One of its neighbors is certainly free because we came from a free neighbor. Every other neighbor is forced with probability $\rho^*$. Thus, the expected number of forced is $(\kappa_{\D}-1)\rho^*$, and the rest are free. 
Substituting Eq.\ \eqref{kappaDDkappaDFarbitrary} into Eq.\ \eqref{XForced1} we get the Mean-Field equation for the states of a forced system,
\begin{equation}
	R(\bar{\x}) = \lambda (1+(\kappa_{\D}-1)(1-\rho^*))   M_2(\bar{\x}) + \lambda (\kappa_{\D}-1)\rho^* M_2(\Delta).
	\label{ForcedMFArb}
\end{equation}
Considering $\kappa_{\D}\gg 1$ and denoting 
\begin{equation} \label{betaD}
	\beta_{\D}=\lambda\kappa_{\D}, 
\end{equation}
we get a similar equation to Eq.\ \eqref{ForcedMFRandomBeta} but for a general degree distribution of the controlled nodes,
\begin{equation}
	R(\bar{\x}) = \beta_{\D}(1-\rho^*)  M_2(\bar{\x}) + \beta_{\D}\rho^*  M_2(\Delta).
	\label{ForcedMFArbitraryBeta}
\end{equation}
This equation allows us to predict the states of the forced system and thus to find the window of sustainability as shown above. \\
Yet, there is a significant difference between Eq.\ \eqref{ForcedMFArbitraryBeta}, and the one of random selection, \eqref{ForcedMFRandomBeta}, because here $\beta_{\D}$ depends on $\rho$ and the moments of $f_k$ (degree distribution of $\F$), Eqs.\ \eqref{betaD} and \eqref{kappaD}. Also, $\rho^*$ depends on the moments of $p_k$ and $f_k$, Eq.\ \eqref{RhoStar}, unlike for a random selection. Thus, the separation of parameters of the topology, $\beta$, and the dynamic intervention, $\rho$ and $\Delta$, does not work in this general selection case.

Finally, we find the quantities of Eq.\ \eqref{ThetaSC5}, to obtain the self-consistent equation for large fluctuations,
\begin{equation}
	\begin{array}{lcl}
		\Pr(k_0) & = & \Pr(k_0|\D) = (p_{k_0} - \rho f_{k_0}) / (1-\rho),
		\\[10pt]
		\Pr(k|k_0) & = & \binom{k_0}{k}(1-\rho^*)^k {\rho^*}^{k_0-k},
		\\[10pt]
		\av{k_{\D\to\D}} & = & \kk_{\D}(1-\rho^*).
 	\end{array}	
	\label{Pk0Pkk0}	
\end{equation}
Plugging these three terms into Eq.\ \eqref{ThetaSC5}, we obtain,
\begin{equation}
	\begin{aligned}
		\Theta & = \frac{1}{\kk_{\D}(1-\rho^*)}  \times
		\\
		& \sum_{k_0} \frac{p_{k_0} - \rho f_{k_0}}{1-\rho}\sum_{k \leq k_0} k \binom{k_0}{k}(1-\rho^*)^k {\rho^*}^{k_0-k} M_2 \left(R^{-1} \left(\lambda k \Theta  + \lambda (k_0-k)  M_2(\Delta) \right) \right) ,
	\end{aligned}
	\label{ThetaSCForcedArb}
\end{equation}
which provides, for given degree distributions of the network and of the forced nodes, the order parameter $\Theta$ as a function of $\lambda$, $\rho$, and $\Delta$. Eq.\ \eqref{ThetaSCForcedArb}, for an arbitrary degree distribution of controlled nodes, replaces Eq.\ \eqref{ThetaSCForcedSol}, which stands for a random selection.


\subsection{Sustainability window}
\label{SecWinArb}

From Eq.\ \eqref{ForcedMFArbitraryBeta}, we can find the conditions under which the system is sustainable, and the limits of this sustainable region, \textit{i.e.}\ the transitions point. 
Since Eq.\ \eqref{ForcedMFArbitraryBeta} has the same shape as Eq.\ \eqref{ForcedMFRandomBeta}, the same condition on $\beta,\rho,\Delta$ for a random selection, which is obtained by Eq.\ \eqref{Maximum}, applies also to $\beta_{\D},\rho^*,\Delta$ for a general selection. Thus, at the transition we get a relation between those three quantities $\beta_{\D},\rho^*,\Delta$, and \textit{e.g.}\ for given $\beta_{\D}$ and $\Delta$, if $\rho^*$ is larger than the value obtained from the transition relation, the system is sustainable. The same holds for $\beta_{\D}$ or $\Delta$. If they are larger than their value at criticality the system is sustainable, see Fig.\ \ref{fig:Selection} for results for cellular dynamics.

However, it is more complicated than the random case, because $\beta_{\D}$ and $\rho^*$ are not independent variables and also not directly controllable, thus we cannot keep one of them fixed, and then change the other until it gets the critical value. For example, increasing $\rho$ enlarges $\rho^*$ but in turn might decrease $\kappa_{\D}$ and as a result also $\beta_{\D}$. Nevertheless, for given $p_k$ and all the properties of the control, $\rho,f_k,\Delta$, we can find the critical interaction strength, $\lambda_c$, because it appears in a simple way only in $\beta_{\D}=\lambda\kappa_{\D}$.

In specific ways of selecting the set $\F$, we can find simplified expressions for $\beta_{\D}$ and $\rho^*$ obtain an explicit condition for sustainability. We shall consider one example of non-random selection, which is selecting a fraction $\rho$ of the nodes with the highest degrees to be controlled.

\subsection{High degree nodes selection}

In this Section, to explore the impact of targeted control aimed to the most influential nodes, we select as the forced set, $\F$, a fraction $\rho$ of the network, comprising of the nodes with the highest degrees.
We aim to find explicit expressions for the terms in Eq.\ \eqref{ForcedMFArbitraryBeta}, $\beta_{\D}$ and $\rho^*$.
We do this for power-law degree distribution which characterizes scale-free networks.
 
\subsubsection{Scale-free networks}

\begin{figure}[h]
	\centering
	\includegraphics[width=0.55\linewidth]{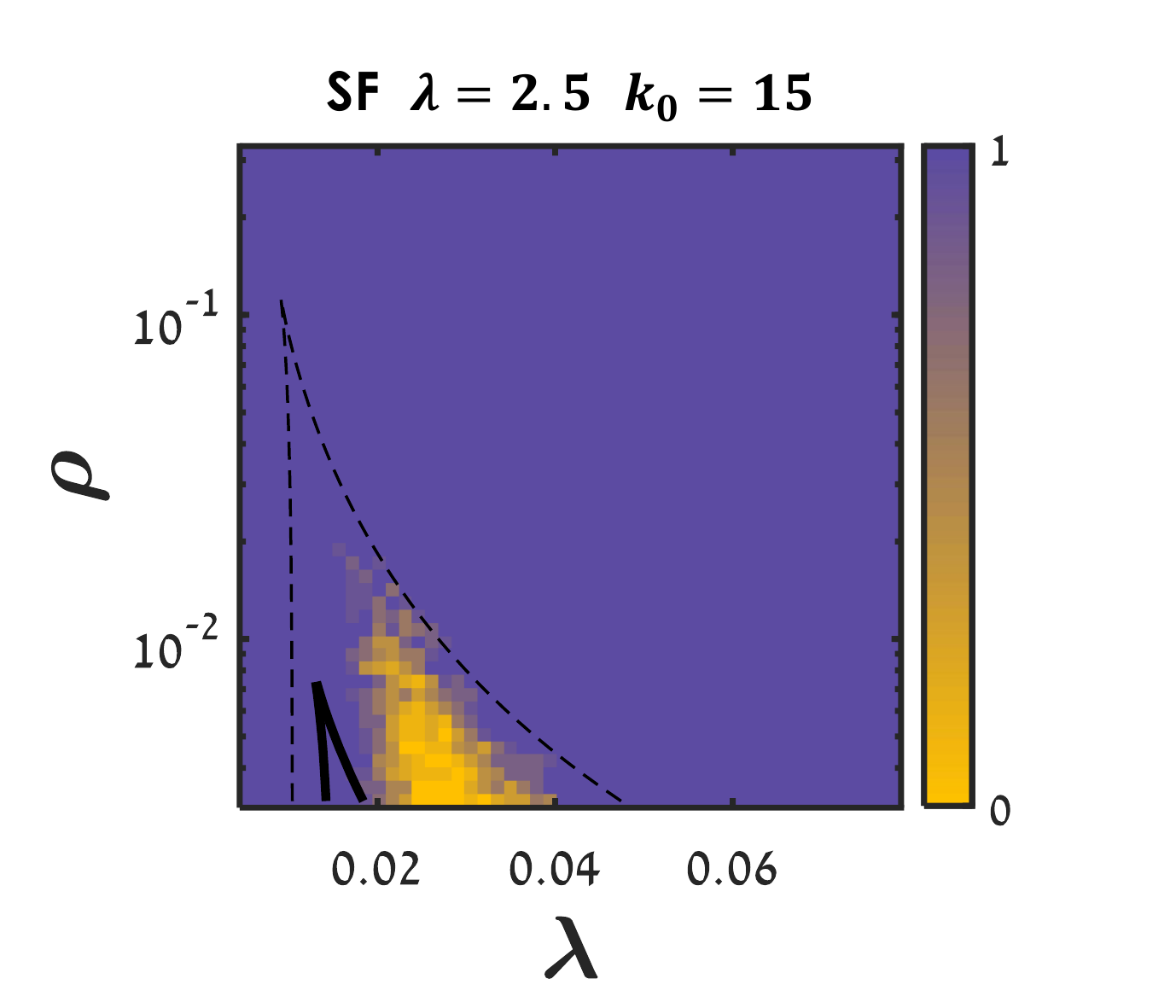}
	\caption{ 
		\textbf{Results for scale-free networks with small exponent $\gamma$.} 
		In contrast to Fig.\ 4a,b in the main text for SF network with $\gamma=3.5$, in which the theory of Eq.\ \eqref{ThetaSCForcedSol} is excellent while the theory of small fluctuations, Eq.\ \eqref{BetaXForced} has a deviation from simulations, here, for SF with $\gamma=2.5$ which has larger degree variation, the theory of Eq.\ \eqref{ThetaSCForcedSol} fails. However, the theory of Eq.\ \eqref{BetaXForced} still works quite well. However, it might be coincidentally since Eq.\ \eqref{BetaXForced} relies upon Eq.\ \eqref{ThetaSCForcedSol} and based on further assumption, so by definition it is not capable of providing better results.
	}
	\label{fig:SF}	
\end{figure}

Here we consider scale-free (SF) networks whose degree distribution allows an extensive analysis. 
Another importance of SF is that the variation of its degrees is large, thus we expect that different selections of the forced nodes would impact significantly.
However, these networks challenge the MF of small fluctuations represented by Eq.\ \eqref{ForcedMFArbitraryBeta}, and the MF might be inaccurate in some cases. Thus, all the following results for SF are useful only where MF is valid also for SF networks. Yet, as we show in Fig.\ \ref{fig:SF}c, the scaling relation can be predicted quite good although quantitatively the theory does not recover well the simulations, see Fig.\ 4b in the main text.
   
Let the degree distribution be 
\begin{equation}
	p_k=Ak^{-\gamma},
\end{equation}
for $k_0\leq k \leq k_f$, and $p_k=0$ otherwise. The minimal degree $k_0$ is set arbitrarily, while the maximal degree $k_f$ is obtained naturally as the \emph{natural cutoff} which is evaluated by $k_f\approx k_0 N^{1/(\gamma-1)}$ \cite{cohen2000prl}. 
We analyze separately three regimes of the exponent $\gamma$, (i) $\gamma>3$ which yields a finite variance of the degree, (ii) $2<\gamma<3$ which yields a divergent variance of the degree, (iii) $\gamma=3$ with a divergent variance as well.

\subsubsection*{ \boldmath $\gamma>3$ }

First we consider the interval of the exponent, $\gamma>3$, such that the second moment $\kkk$ is finite.
For such distribution, the moments are
\begin{equation}
	\begin{aligned}
		\kk & \approx \dfrac{\int_{k_0}^{k_f} k k^{-\gamma}\, {\rm d}k}{\int_{k_0}^{k_f} k^{-\gamma}\, {\rm d}k}   =  \dfrac{\gamma-1}{\gamma-2}\, \dfrac{k_f^{2-\gamma}-k_0^{2-\gamma}}{k_f^{1-\gamma}-k_0^{1-\gamma}}
		 \approx  \dfrac{\gamma-1}{\gamma-2}\, k_0 ,
		\\[10pt]
		\kkk & \approx \dfrac{\int_{k_0}^{k_f} k^2 k^{-\gamma}\, {\rm d}k}{\int_{k_0}^{k_f} k^{-\gamma}\, {\rm d}k}   = 
		\dfrac{\gamma-1}{\gamma-3}\,\dfrac{k_f^{3-\gamma}-k_0^{3-\gamma}}{k_f^{1-\gamma}-k_0^{1-\gamma}}
 		\approx  \dfrac{\gamma-1}{\gamma-3}\,k_0^2 ,
		\\[10pt]
		\kappa & \approx \dfrac{\gamma-2}{\gamma-3}\,k_0 ,
	\end{aligned}
\label{DegreesSF}
\end{equation}
where we assumed $k_0\ll k_f$.

Next, we find $k_s$, which is the edge degree separating between the forced and free nodes. Above $k_s$, all degrees of nodes in $\F$ are located, and below, all nodes are free. Thus, 
\begin{gather}
	\rho \approx \dfrac{\int_{k_s}^{k_f} k^{-\gamma}\, {\rm d}k}{\int_{k_0}^{k_f} k^{-\gamma}\, {\rm d}k  }  = \dfrac{k_f^{1-\gamma}-k_s^{1-\gamma}}{k_f^{1-\gamma}-k_0^{1-\gamma}}
	\approx \bigg(\frac{k_0}{k_s}\bigg)^{\gamma-1} ,
	\label{RhoSF}
\end{gather}
where we assumed $k_s\ll k_f$. Consequently,
\begin{equation}
	k_s \approx k_0 \rho ^ {-\frac{1}{\gamma-1}}.
\end{equation}
Now, over the forced nodes, the average degree is
\begin{equation}
	\kk_{\F} \approx \dfrac{ \int_{k_s}^{k_f} k k^{-\gamma}\, {\rm d}k }{\int_{k_s}^{k_f} k^{-\gamma}\, {\rm d}k}  
	= \frac{\gamma-1}{\gamma-2} \dfrac{k_f^{2-\gamma}-k_s^{2-\gamma}}{k_f^{1-\gamma}-k_s^{1-\gamma}}
	\approx \frac{\gamma-1}{\gamma-2} k_s	.
	\label{kFgamma}
\end{equation}
Using Eqs.\ \eqref{DegreesSF}-\eqref{kFgamma}, we get
\begin{equation} \label{kFSF}
	\kk_{\F} \approx \kk \frac{k_s}{k_0} \approx \kk \rho^{-\frac{1}{\gamma-1}}.
\end{equation}
Thus, based on Eqs.\ \eqref{RhoStar} and \eqref{kFSF}, we obtain
\begin{equation}
	\rho^* = \rho \kk_{\F} /\kk \approx \rho^{\frac{\gamma-2}{\gamma-1}}.
	\label{RhoStarHigh}
\end{equation}
Next, we move to find $\kappa_{\D}$ which appears in Eqs.\ \eqref{ForcedMFArbitraryBeta}, \eqref{betaD} and \eqref{kappaD}. The second moment of degree over the free nodes is
\begin{equation} \label{k2Dgamma}
	\kkk_{\D} \approx \dfrac{\int_{k_0}^{k_s} k^2 k^{-\gamma}\, {\rm d}k }{\int_{k_0}^{k_s} k^{-\gamma}\, {\rm d}k}  
	=  \frac{\gamma-1}{\gamma-3} \dfrac{k_s^{3-\gamma}-k_0^{3-\gamma}}{k_s^{1-\gamma}-k_0^{1-\gamma}}
	\approx \frac{\gamma-1}{\gamma-3} k_0^2 
	\approx \kkk ,
\end{equation}
where we assumed $k_0\ll k_s$. The first moment is
\begin{equation} \label{kDgamma}
	\kk_{\D} \approx \dfrac{\int_{k_0}^{k_s} k k^{-\gamma}\, {\rm d}k }{\int_{k_0}^{k_s} k^{-\gamma}\, {\rm d}k}  
	=  \frac{\gamma-1}{\gamma-2} \dfrac{k_s^{2-\gamma}-k_0^{2-\gamma}}{k_s^{1-\gamma}-k_0^{1-\gamma}}
	\approx \frac{\gamma-1}{\gamma-2} k_0 
	\approx \kk .
\end{equation}

Therefore,
\begin{equation} \label{kappaDSF}
	\kappa_{\D} = \frac{\kkk_{\D} }{\kk_{\D} }
	\approx \kappa,
\end{equation}
and thus,
\begin{equation} \label{betaDSF}
	\beta_{\D}
	=\lambda \kappa_{\D} 
	\approx \beta .
\end{equation}

Substituting Eqs.\ \eqref{betaDSF} and \eqref{RhoStarHigh} into Eq.\ \eqref{ForcedMFArbitraryBeta}, we get the equation for system states when forcing the highest-degree nodes of SF with $\gamma>3$. Since for this case the expressions of $\beta_{\D}$ and $\rho^*$ depend only on $\beta$ and $\rho$ independently, we can just use the relation for \emph{random selection}, Section \ref{SecWin}, and replace the pair $\beta$, $\rho$, by $\beta_{\D}$ and $\rho^*$ that we obtained here.

If the relation $\rho_c(\beta)$ for random selection has a power-law scaling as in cellular dynamics, Eq.\ \eqref{RhocMMScaling},
\begin{equation}
	\rho_c \sim \beta^{-\theta},
\end{equation}
then, for high degrees selection, in SF network with $\gamma>3$,
\begin{equation}
	\rho_c^* \sim \beta_{\D}^{-\theta}.
\end{equation}
Substituting into this Eqs.\ \eqref{RhoStarHigh} and \eqref{betaDSF}, we obtain
\begin{equation}
	\rho_c^{\frac{\gamma-2}{\gamma-1}} \sim \beta^{-\theta},
\end{equation}
yielding the scaling,
\begin{equation} \label{ScalingRhoBetaArb}
	\rho_c \sim \beta^{-\theta \frac{\gamma-1}{\gamma-2}}.
\end{equation}
See Section \ref{SecCell} and Fig.\ \ref{fig:Selection}c for cellular dynamics.

\subsubsection*{\boldmath $2<\gamma<3$}

For degree distribution $p_k=Ak^{-\gamma}$ with $2<\gamma<3$ for $k_0\leq k \leq k_f$, the second moments $\kkk$, $\kkk_{\F}$ and $\kkk_{\D}$ diverge, and their leading term involves $k_{f}$ or $k_s$. The first moments, however, stay the same. Thus, $\rho^*$ remains the same as in Eq.\ \eqref{RhoStarHigh}, while $\beta_{\D}$ should be recalculated. Hence, we go back to Eq.\ \eqref{DegreesSF} and recalculate, for the leading term,
\begin{equation} \label{k2gamma23}
		\kkk  \approx \dfrac{\int_{k_0}^{k_f} k^2 k^{-\gamma}\, {\rm d}k}{\int_{k_0}^{k_f} k^{-\gamma}\, {\rm d}k}   = \frac{\gamma-1}{\gamma-3} \frac{k_f^{3-\gamma}-k_0^{3-\gamma}}{k_f^{1-\gamma}-k_0^{1-\gamma}}
		\approx \frac{\gamma-1}{3-\gamma} k_f^{3-\gamma}k_0^{\gamma-1}  .
\end{equation}
In similar way,
\begin{equation} \label{k2Dgamma23}
			\kkk_{\D}
			\approx \frac{\gamma-1}{3-\gamma} k_s^{3-\gamma}k_0^{\gamma-1} .
\end{equation}
Using Eq.\ \eqref{k2gamma23}, Eq.\ \eqref{k2Dgamma23} becomes
\begin{equation} \label{k2Dgamma232}
	\kkk_{\D} 
	\approx \kkk \Big( \frac{k_s}{k_f} \Big)^{3-\gamma} .
\end{equation}
The first moment is still $\kk_{\D}\approx\kk$ as in Eq.\ \eqref{kDgamma}.
Therefore,
\begin{equation} \label{kappaDSF23}
	\kappa_{\D} = \frac{\kkk_{\D} }{\kk_{\D} }
	\approx \kappa \Big( \frac{k_s}{k_f} \Big)^{3-\gamma} 
	\approx \kappa \Big( \frac{k_s}{k_0} \Big)^{3-\gamma} \Big( \frac{k_0}{k_f} \Big)^{3-\gamma} ,
\end{equation}
and using Eq.\ \eqref{RhoSF} we obtain 
\begin{equation} \label{kappaDSF232}
	\kappa_{\D} 
	\approx \kappa  \Big( \frac{k_0}{k_f} \Big)^{3-\gamma} \rho^{-\frac{3-\gamma}{\gamma-1}} .
\end{equation}
Using the natural cutoff for $k_f$, we know that \cite{cohen2000prl}
\begin{equation} \label{cutoff}
	k_f	\approx k_0 N^{\frac{1}{\gamma-1}} .
\end{equation}
Therefore Eq.\ \eqref{kappaDSF232} turns to
\begin{equation} \label{kappaDSF233}
	\kappa_{\D} 
	\approx \kappa  N^{-\frac{3-\gamma}{\gamma-1}} \rho^{-\frac{3-\gamma}{\gamma-1}} .
\end{equation}
resulting in
\begin{equation} \label{betaDSF23}
	\beta_{\D} = \lambda \kappa_{\D}
	\approx \beta N^{-\frac{3-\gamma}{\gamma-1}} \rho^{-\frac{3-\gamma}{\gamma-1}} .
\end{equation}

Here, differently from the above case of $\gamma>3$, $\beta_{\D}$ is not independent variable but depends on both $\beta$ and $\rho$. Therefore, for a given $\beta$, increasing $\rho$ changes both $\rho^*$ and $\beta_{\D}$, thus there is no guarantee that $\rho_c$ will be exactly at the value according to the relation $\rho^*_c(\beta_{\D})$ derived from the relation $\rho_c(\beta)$ of random selection.

\subsubsection*{\boldmath $\gamma=3$}

Let us find the second moments and $\kappa_{\D}$ for this case.
\begin{equation} \label{k2gamma3}
	\kkk  \approx \dfrac{\int_{k_0}^{k_f} k^2 k^{-3}\, {\rm d}k}{\int_{k_0}^{k_f} k^{-3}\, {\rm d}k}  = 2 \frac{\ln k_f - \ln k_0 }{k_0^{-2}-k_f^{-2}}
	\approx 2  k_0^2 \ln (k_f / k_0).
\end{equation}
In a similar way,
\begin{equation} \label{k2Dgamma3}
	\kkk_{\D}
	\approx 2  k_0^2 \ln (k_s / k_0).
\end{equation}
Using Eqs.\ \eqref{k2gamma3} and \eqref{RhoSF}, Eq.\ \eqref{k2Dgamma3} becomes
\begin{equation} \label{k2Dgamma32}
	\kkk_{\D} 
	\approx \kkk \frac{1}{\ln(k_f/k_0)} \frac{1}{2} \ln \Big(\frac{1}{\rho} \Big).
\end{equation}
The first moment is still $\kk_{\D}\approx\kk$ as in Eq.\ \eqref{kDgamma}.
Therefore,
\begin{equation} \label{kappaDSF3}
	\kappa_{\D} = \frac{\kkk_{\D} }{\kk_{\D} }
	\approx \kappa \frac{1}{\ln(k_f/k_0)} \frac{1}{2} \ln \Big(\frac{1}{\rho} \Big) .
\end{equation}
Using the natural cutoff, $k_f \approx k_0 N ^ {1/2}$, we get
\begin{equation} \label{kappaDSF32}
	\kappa_{\D}	\approx \kappa \frac{1}{\ln N} \ln \Big(\frac{1}{\rho} \Big) ,
\end{equation}
resulting in
\begin{equation} \label{betaDSF3}
	\beta_{\D} = \lambda \kappa_{\D}
	\approx \beta \frac{1}{\ln N} \ln \Big(\frac{1}{\rho} \Big) .
\end{equation}

Also here, similarly to $\gamma<3$, $\beta_{\D}$ depends both on $\beta$ and on $\rho$.

\clearpage

\section{Distance-based method for the limit \boldmath$\rho\to0$}
\label{SecDistanceBMF}

In this section we introduce a method to find the required fraction $\rho_c$ for sustaining a network, also in the limit of $\rho\to0$ in which our above MF theory, Eqs.\ \eqref{ForcedMFrandom} and \eqref{ThetaSCForcedSol}, fails. The above MF predicts that if $\rho\to0$, then the phase diagram of the system converges to that of a free system which is obtained by substituting $\rho=0$. Namely, a single node which is a zero fraction cannot make a global change. However, simulation results and theory \cite{sanhedrai2022reviving} show otherwise. Under certain conditions, controlling a single node makes a system sustainable. Therefore, we generalize here the method in Ref.\ \cite{sanhedrai2022reviving} developed for controlling only a single node, and develop a method that covers both limits of a few controlled nodes and of an extensive fraction of controlled nodes.

We recognize that the above MF does not cover the case of a microscopic number of controlled nodes since the forced nodes break the uniformity and homogeneity of nodes along the network and divide the nodes to classes according to their distance from the forced nodes. Ref.\ \cite{sanhedrai2022reviving} used a shells-based mean-field around the single controlled node. Here we do a similar assumption for many forced nodes. We perform a distance-based mean-field where we assume that nodes having the same distance from the closest controlled node behave alike.

\begin{figure}[h]
	\centering
	\includegraphics[width=0.99\linewidth]{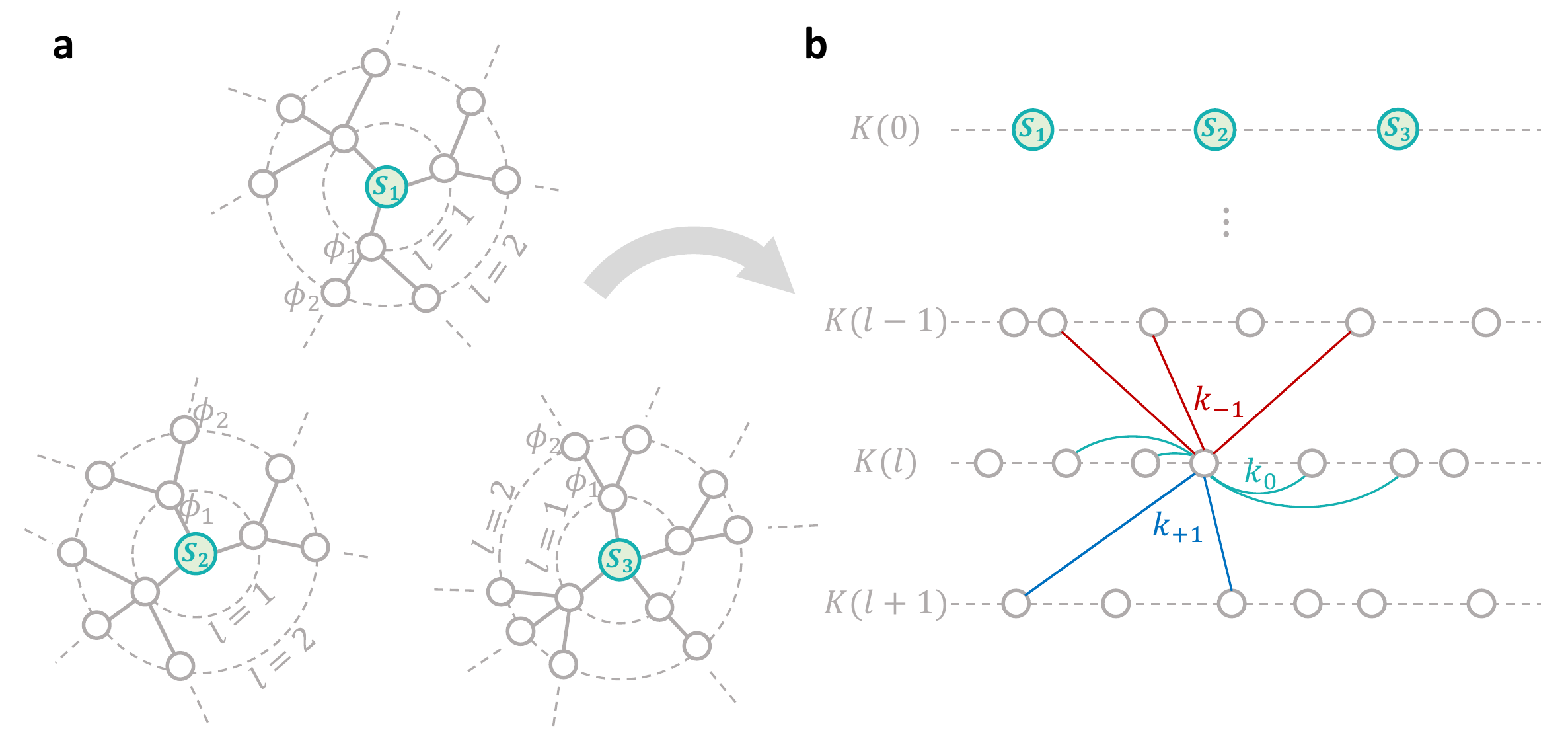}
	\caption{ \footnotesize
		\textbf{Distance-Based MF for covering both limits of $\rho>0$ and $\rho\to0$.} 
		(a) Controlling multiple nodes $s_1,s_2,s_3,...\in \F$, separates the network into shells around each one of the controlled nodes. However, these ''circles'' meet at some finite $l$ and shells become blended. Thus, we decide to separate the network according to the distance of each node from the closest forced node $s_i\in\F$. The average activity in distance $l$ from the closest controlled node is denoted by $\phi_l$, Eq.\ \eqref{phil}.
		(b) We denote $K(l)$, Eq.\ \eqref{Kl}, as the set of all the nodes in distance $l$ from the closest forced node. Hence, the set $K(0)$ contains all the forced nodes, \textit{i.e.}\ the complete set $\F$. $K(l-1)$ contains all nodes in distance $l-1$ from their closest forced node $s_i\in\F$. The same are $K(l)$ and $K(l+1)$. To find a relation between consecutive distances we define the number of neighbors of a node in the same distance, $k_0$, in a distance lower in one, $k_{-1}$, and in a distance larger in one, $k_{+1}$, as defined in Eq.\ \eqref{kni}. For instance, the node presented in this demonstration has $k_{-1}=3$, $k_0=4$, and $k_{+1}=2$. Based on this separation of the network into distances, we obtain a mean field approximation, Eq.\ \eqref{Reccurence} that gives a good prediction for $\rho_c$ for both limits of $\rho_c>0$ and $\rho_c\to0$ as shown in Fig.\ 3l in the main text.
	}
	\label{fig:shells}	
\end{figure}

Thus, we denote $K(l)$ as the set of all nodes in distance $l$ from the closest controlled node,
\begin{equation}
	K(l) = \bigg\{ i \  \Big| \min_{j\in\F} d(i,j) = l \bigg\},
	\label{Kl}
\end{equation}
where $d(i,j)$ is the distance between nodes $i$ and $j$, and $\F$ is the set of the forced nodes.
Next, we denote $\phi_l$ as the average activity in relaxation ($x_i^*$) over all the nodes in distance $l$ from the closest controlled node, 
\begin{equation}
	\phi_l = \frac{1}{|K(l)|} \sum_{i\in K(l)} x_i^*,
	\label{phil}
\end{equation}
where $x_i^*=x_i(t\to\infty)$ is the relaxation activity of node $i$, and $|K(l)|$ refers to the size of the set $K(l)$ defined in \eqref{Kl}.
Next, we seek for an equation connecting between the average activities $\phi_l$ of different distances $l$'s.
Hence, we denote the number of neighbors in $K(l+n)$ of a node $i\in K(l)$ as $k_{n,i}$, that is,
\begin{equation}
	k_{n,i} = \sum_{j\in K(l+n)} \m Aij .
	\label{kni}
\end{equation}
We notice that a node in $K(l)$, has neighbors only in $K(l-1)$, $K(l)$, or $K(l+1)$. Thus $k_{n,i}$ is zero for any $n$ unless $n=-1,0,1$.
See Fig.\ \ref{fig:shells} for a demonstration. 
Since we seek for the average behavior in distance $l$ from the controlled nodes, we denote the average of $k_{n,i}$ over $K(l)$ by $k_n(l)$,
\begin{equation}
	k_{n}(l) = \frac{1}{|K(l)|}\sum_{i\in K(l)} k_{n,i}.
	\label{knl}
\end{equation}
This quantity, $k_n(l)$, depends both on $l$ but also on $\rho$, thus we denote it by $k_n(l,\rho)$. For instance, if $l=l_{\rm max}$, \textit{i.e.}\ the maximal distance from a forced node, then $k_{+1}(l_{\rm max},\rho)=0$ since there are no neighbors in distance $l+1$. In addition, the value of $l_{\rm max}$ depends on the value of $\rho$. This implies that $k_n(l,\rho)$ depends on both $l$ and $\rho$. 

Suppose we have for a given network and a given set of controlled nodes $\F$ the quantity $k_n(l,\rho)$. Let us focus on a node in distance $l$ from the closest forced node. The dynamics of this node, following Eq.\ \eqref{Dynamics}, is
\begin{equation}
	\dod{x_i}{t} = M_0(x_i) + \lambda M_1(x_i) \sum_{n=-1}^{+1} \, \sum_{j \in K(l+n)} \m Aij  M_2(x_j),
\end{equation}
where $i\in K(l)$.
In relaxation the derivative vanishes, and we obtain
\begin{equation}
	 R(x_i^*) = \lambda \sum_{n=-1}^{+1} \, \sum_{j \in K(l+n)} \m Aij  M_2(x_j^*),
\end{equation}
where $R(x)=-M_0(x)/M_1(x)$. Next, we apply our mean-field approximation where we assume that inside $K(l)$ nodes are similar, thus we replace $x_j^*\in K(l)$ by the average $\phi_l$ defined in \eqref{phil}. Thus we obtain
\begin{equation}
	R(x_i^*) = \lambda \sum_{n=-1}^{+1} M_2(\phi_{l+n}) \sum_{j \in K(l+n)} \m Aij .
\end{equation}
Using Eq.\ \eqref{kni} it turns to
\begin{equation}
	R(x_i^*) = \lambda \sum_{n=-1}^{+1} k_{n,i} M_2(\phi_{l+n}) .
\end{equation}
Averaging the nodes in $i\in K(l)$, we replace $x_i^*$ by $\phi_l$, and $k_{n,i}$ by the averages $k_n(l,\rho)$ defined in \eqref{knl}, to obtain
\begin{equation}
	R(\phi_l) = \lambda \sum_{n=-1}^{+1} k_n(l,\rho) M_2(\phi_{l+n}) .
\end{equation}
To summarize, we get a second order recurrence relation for $\phi_l$,
\begin{equation}
	\begin{cases}
		\phi_0=\Delta,
		\\[10pt]
		R(\phi_l) = \lambda k_{-1}(l,\rho) M_2(\phi_{l-1}) + \lambda k_{0}(l,\rho) M_2(\phi_{l}) + \lambda k_{+1}(l,\rho) M_2(\phi_{l+1}), & 1 \leq l \leq l_{\rm max} ,
	\end{cases}	
	\label{Reccurence}
\end{equation}
where $l_{\rm max}$ is the maximal distance from the controlled nodes all over the network. The value $\phi_{l_{\rm max}+1}$ is not defined since there is no nodes in $K(l_{\rm max}+1)$. However, since $k_{+1}(l_{\rm max})=0$, the variable $\phi_{l_{\rm max}+1}$ does not really appear in the last equation for $\phi_{l_{\rm max}}$, thus it is not a problem.
The initial condition in Eq.\ \eqref{Reccurence} is $\phi_0=\Delta$ since the nodes in distance $0$ from the forced nodes are the forced nodes themselves, and their activity is forced to be $\Delta$. 
To conclude, Eq.\ \eqref{Reccurence} provides $l_{\rm max}$ equations for $l_{\rm max}$ variables, and as such, it is solvable numerically.

Once we have the solution for $\phi_l$, we have to determine whether the system has two solutions, inactive state $\x_0$ and active state $\x_1$, or only $\x_1$ exists, and in other words, whether $\x_0$ exists. If and only if $\x_0$ does not exist then the system is sustainable. For the variable $\phi_l$ two solutions are possible, convergence to $\bar{\x}_1$ of the free system for large $l$, or convergence to $\bar{\x}_0$ of the free system for large $l$. Note that indeed if we seek for a fixed point of the recurrence relation, \eqref{Reccurence}, by demanding $\phi_{l+1}=\phi_{l}=\phi$, we get the equation $R(\phi)=\lambda \kappa M_2(\phi)$ as Eq.\ \eqref{BetaXFree}, and thus the two stable fixed points are $\bar{\x}_1$ and $\bar{\x}_0$. However, the initial condition with high value $\phi_0=\Delta$ might cause that the solution of a convergence to $\bar{\x}_0$ for large $l$ does not exist.

We found numerically if the solution of convergence to $\bar{\x}_0$ exists or only the solution of convergence to $\bar{\x}_1$ exists, to find the critical $\lambda_c$ needed for sustaining a given network for each $\rho$. By this, we plotted the dashed lines representing the results of this theory in Fig.\ 3l in the main text. Note that we used numerical calculations not only for solving the set of equations \eqref{Reccurence}, but also to obtain the expression $k_n(l,\rho)$ defined in \eqref{knl} that appears in \eqref{Reccurence}. We do not have $k_n(l,\rho)$ theoretically, and thus we measure it on many constructed networks and average the results. For solving Eq.\ \eqref{Reccurence} we used the method of adding the term $\dif \phi_l / \dif t$ and letting the system reach a relaxation since then $\dif \phi_l / \dif t = 0$. 

\clearpage

\section{Dynamical models}
\label{SecDynamicModels}

To demonstrate our framework, we examined the sustainability of three relevant dynamic systems. Below we detail the analytical treatment of each of these systems, starting from the \textit{free system}, in which we examine the states of the system without forcing $\Delta$, then treating the \textit{forced system}, in which we control the activity of a fraction of the system.

\subsection{Cellular dynamics}
\label{SecCell}

We consider gene-regulatory dynamics, as captured by the Michaelis-Menten model \cite{Alon2006,Karlebach2008}, for which (\ref{Dynamics}) takes the form

\begin{equation}
\dod{x_i}{t} = -Bx_i^a + \lambda \sum_{j = 1}^N \m Aij \dfrac{x_j^h}{1 + x_j^h}.
\label{MMexample}
\end{equation} 

\noindent
Under this framework $M_0(x_i) = -B x_i^a$, describing degradation ($a = 1$), dimerization ($a = 2$) or a more complex bio-chemical depletion process (fractional $a$), occurring at a rate $B$ \cite{Barzel2011}. For simplicity, in our simulations we set $B = 1$. The activation interaction is captured by a Hill function of the form $M_1(x_i) = 1$, $M_2(x_j) = x_j^h/(1 + x_j^h)$, a \textit{switch-like} function that saturates to $M_2(x_j) \rightarrow 1$ for large $x_j$, representing $j$'s positive, albeit bounded, contribution to $i$'s activity $x_i(t)$. 

\subsubsection{Free system}

First we seek the natural fixed-points of (\ref{MMexample}), by mapping it to the one dimensional space of $\bar{\x}$ via Eq. \eqref{RXBetaFree}. The fixed-points follow

\begin{equation} 
	\bar{\x}_0 = 0,
	\label{betaXMM0}
\end{equation}

the inactive state, and

\begin{equation}
	\beta = \bar{\x}^a \left( 1 + \bar{\x}^{-h} \right),
	\label{betaXMM1}
\end{equation}

whose solutions provide the potentially active and intermediate states. For $a = 1,\ h = 2$, the system we examine in the main text, Eqs.\ (\ref{betaXMM0}) and (\ref{betaXMM1}) provide the three solutions shown in Fig.\ \ref{fig:Cellular}a in the main text (see also \textbf{Box 1}):\ an always stable $\bar{\x}_0$ (red), and a stable $\bar{\x}_1$ (green) for $\beta > \beta_c$. The basins of attraction of $\bar{\x}_0$ and $\bar{\x}_1$ are separated by the intermediate unstable state $\bar{\x}_2$ (gray dashed line).

To obtain the critical point $\beta_c$, we find the transition point where $\beta$ versus $\bar{\x}$ in Eq.\ \eqref{betaXMM1} has an extremum as demanded according to Eq.\ \eqref{MaximumFree}, yielding
\begin{equation}
a \bar{\x}^{a - 1} \left( 1 + \bar{\x}^{-h} \right) - h \bar{\x}^{a} \bar{\x}^{-h-1} = 0.
\end{equation}  
The solution of this equation is
\begin{equation} \label{XcMM}
	\bar{\x}_c = \left( \frac{h}{a}-1 \right)^{1/h} .
\end{equation}
Substituting this in Eq.\ (\ref{betaXMM1}) we arrive at the solution for the critical value of $\beta$,
\begin{equation} \label{BetacMM}
	\beta_c = \frac{h}{a} \left( \frac{h}{a}-1 \right)^{a/h-1},
\end{equation}
capturing the bifurcation in Fig.\ \ref{fig:Cellular}a, where the active state (green) emerges as a stable fixed-point at $\beta \ge \beta_c$.
Setting $a=1, \ h=2$, the parameters used in the main-text simulations we obtain
\begin{equation}
\begin{aligned}
&\bar{\x}_c = 1
\\
&\beta_c = 2,
\end{aligned}
\end{equation}
precisely the transitions observed in Fig.\ \ref{fig:Cellular}a. Therefore, the free system exhibits an inactive phase for $\beta < 2$, and a bi-stable regime (gray shaded) for $\beta \ge 2$, where both $\bar{\x}_0$ and $\bar{\x}_1$ are potentially stable. Thus, in all the range $\beta\ge2$ the system is regarded \emph{unsustainable} since although there exists an active state $\x_1$, yet the existence of inactive state $\x_0$ puts the system in a danger of falling into dysfunctionality. 

\subsubsection{Forced system}

To examine the behavior of our cellular dynamics (\ref{MMexample}) under sustaining we seek to construct the forced system states according to Eq.\ \eqref{BetaXForced},
\begin{equation}
	\beta = \frac{\bar{\x}^a}{(1-\rho)/(1+\bar{\x}^{-h})+\rho/(1+\Delta^{-h})}.
	\label{BetaXForcedMM}
\end{equation}
This formula generates a new phase diagram for a forced system shown in Fig.\ \ref{fig:Cellular}d (thick curve) for $\rho=0.03$ and $\Delta=5$, exhibiting an s-shape diagram which has now also a new active regime (blue shade), in which only $\x_1$ exists. This regime is regarded as the \emph{sustainable phase}. Note that the simulation results (symbols) agree well with theory. The simulations are for ER network with $\kappa=40$ and $N=10^4$. The value of $\lambda$ varies. 

\subsubsection{Sustaining a network}

\begin{figure}[h]
	\centering
	\includegraphics[width=0.6\linewidth]{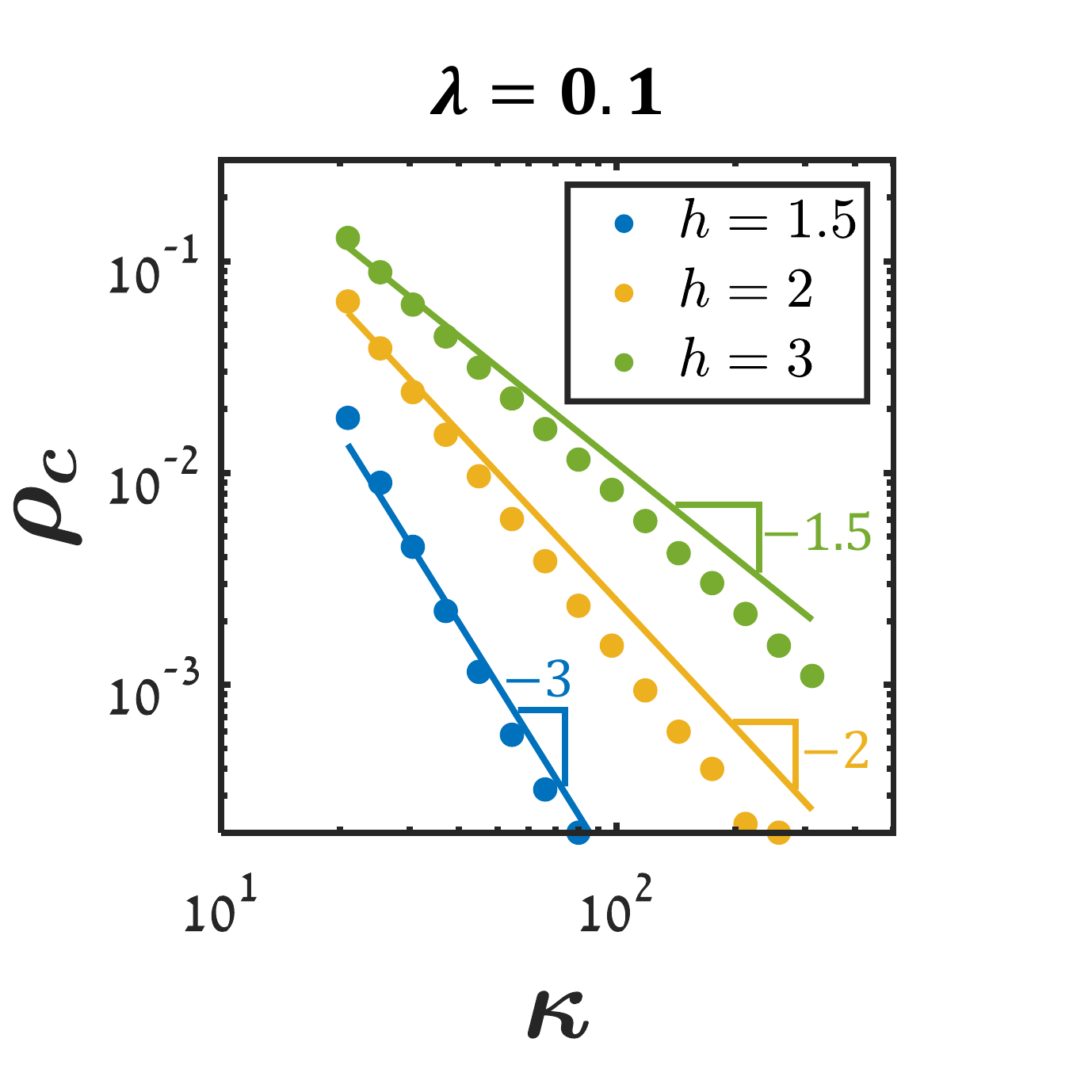}
	\caption{ 
		\textbf{The scaling of $\rho_c$ and $\kappa$ for different exponents of cellular dynamics.} 
		For random selection of controlled nodes, we show how the exponents of dynamics impact the critical fraction for sustaining the network. We set $a=1$ and change $h$. Symbols represent simulations, continuous lines are the slopes obtained from Eq.\ \eqref{RhocMMScaling}. The network is ER with size of $N=10^4$ and the results were averaged over 10 realizations. The interaction strength was set to be $\lambda=0.1$. 
	}
	\label{fig:Expah}	
\end{figure}

For finding $\beta_c$, above which the forced system is sustainable, according to Eq.\ (\ref{Maximum}), we take the derivative of $\beta$ in Eq.\ \eqref{BetaXForcedMM} with respect to $\bar{\x}$ to be zero as Eq.\ \eqref{Maximum} demands, yielding
\begin{equation}
	\begin{aligned}		
		 a\bar{\x}^{a-1}\bigg(\frac{1-\rho}{1+\bar{\x}^{-h}}+\frac{\rho}{1+\Delta^{-h}}\bigg)
		+ \bar{\x}^{a}\frac{(1-\rho)(-h)\bar{\x}^{-h-1}}{(1+\bar{\x}^{-h})^2} = 0 .
	\end{aligned}
\label{DerBetaMM}
\end{equation}
Denoting $u=1+\bar{\x}^{-h}$, we obtain,
\begin{equation} \label{MMTwoSols}
			\frac{a(1-\rho)}{u}+\frac{a\rho}{1+\Delta^{-h}} -
			\frac{h(1-\rho)(u-1)}{u^2} = 0 .
\end{equation}
Arranging terms gives a solvable quadratic equation,
\begin{equation} 
	\frac{a\rho}{1+\Delta^{-h}}u^2 - (h-a)(1-\rho)u  +h(1-\rho) = 0,
\end{equation}
whose solutions are given by
\begin{equation} 
	u = \frac{(h-a)(1-\rho) \pm \sqrt{(h-a)^2(1-\rho)^2-4ah\rho(1-\rho)/(1+\Delta^{-h})}}{2a\rho/(1+\Delta^{-h})} .
\end{equation}
Taking the solution which gives the smaller $\bar{\x}_c$ for the transition to the sustainable regime (see Fig.\ \ref{fig:Cellular}d), we get
\begin{equation} \label{MMRightSol}
	\bar{\x}_c^{-h} = -1 + \frac{(h-a)(1-\rho) - \sqrt{(h-a)^2(1-\rho)^2-4ah\rho(1-\rho)/(1+\Delta^{-h})}}{2a\rho/(1+\Delta^{-h})} .
\end{equation}
Substituting this in Eq.\ \eqref{BetaXForcedMM} provides $\beta_c(\rho,\Delta)$,
yields a complicated expression. However, let us expand this in the limit of small $\rho$ and as a result small $\bar{\x}_c$ as can be seen in Fig.\ \ref{fig:Cellular}d, to get the scaling of $\beta_c$ and $\rho$ in the limit of small $\rho$. To this end, we go back to Eq.\ \eqref{DerBetaMM} to obtain for the lead order
\begin{equation}
	\bar{\x}_c^{h} \approx \frac{\rho}{1+\Delta^{-h}} \frac{1}{(h-a)(1-\rho)}
	\approx \frac{\rho}{1+\Delta^{-h}} \frac{1}{(h-a)}
	\sim \rho ,
\end{equation}
and substituting this in Eq.\ \eqref{BetaXForcedMM}, we finally obtain
\begin{equation} \label{BetacMMScaling}
	\beta_c \approx \frac{\bar{\x}_c^a}{\bar{\x}_c^{h}+\rho/(1+\Delta^{-h})} \sim \rho^{-\frac{h-a}{h}},
\end{equation}
which provides the scaling between the connectivity $\beta$ and the fraction of control $\rho$ at the transition from the unsustainable phase to the sustainable phase for small $\rho$ in cellular dynamics.
The inverse relation gives for a given $\beta$ the critical required fraction $\rho_c$ for sustaining a network,
\begin{equation} \label{RhocMMScaling}
	\rho_c \sim \beta^{-\frac{h}{h-a}}.
\end{equation}
In Fig.\ 3k in the main text we present the results for $a=1,\ h=2$, thus we observe the scaling $\rho_c \sim \beta ^ {-2} = (\lambda\kappa)^{-2}$. In Fig.\ \ref{fig:Expah} we show results also for different values of $a,h$ resulting in different exponents and a good agreement between simulations and theory. 

Notice that as we discussed above in Section \ref{SecDistanceBMF}, that the limits of small $\rho$ and small $\av{k}$ challenge our mean-field that provides the scaling of \eqref{RhocMMScaling} which is valid only for small $\rho$. Thus, even though the simulation results in Fig.\ 3k in the main text show the predicted exponent because of the large degrees, in contrary, the results in Fig.\ 3l show a deviation from the predicted scaling because the MF assumption is broken due to the small $\rho$ and small average degrees.

\subsubsection{Tricritical point}
Beyond some point $(\beta_0,\rho_0)$ the unsustainable phase vanishes, and the transition between the high-active state and the low-active state becomes continuous, see Fig.\ 3d and Fig.\ 3g,h in the main text. This happens when the two transition points on the edges of the unsustainable region merge. In order for this to be fulfilled, there should be only a single solution to Eq.\ \eqref{MMTwoSols}, therefore the term inside the square root has to equal zero. Thus we demand
\begin{gather}
	(h-a)^2(1-\rho)^2 - \frac{4ah\rho(1-\rho)}{1+\Delta^{-h}} = 0 ,
\end{gather}
yielding 
\begin{gather}
	\rho_0 = \frac{1}{1 + \frac{4ah}{(h-a)^2} \frac{1}{1+\Delta^{-h}}} ,
\end{gather}
which is the critical value of $\rho$ above which the system is fully sustainable because the bi-stable regime disappears.
Plugging this in Eqs.\ \eqref{MMRightSol} and \eqref{BetaXForcedMM} provides the value $\beta_c$ of the critical point.

In our simulations in Fig.\ 3 in the main text we set $a=1$, $h=2$, thus we get $\rho_c=1/(1+8/(1+\Delta^{-h}))$, and setting $\Delta = 5$ gives $\rho_c \approx 1/9$, while using $\Delta=1$ yields $\rho_c=1/5$. In general, for high $\Delta$, we get $\rho_c \approx (h-a)^2/(h+a)^2$.

\subsubsection{Sustainability for high-degree nodes selection}

As obtained above, Eq.\ \eqref{ForcedMFArbitraryBeta}, when the forced nodes have an arbitrary degree distribution, and they are spread randomly across the network, then $\beta_{\D}$ (Eq.\ \eqref{betaD}) just replaces $\beta$ while $\rho^*$ (Eq.\ \eqref{RhoStar}) replaces $\rho$. 

\begin{figure}[ht]
	\centering
	\includegraphics[width=0.99\linewidth]{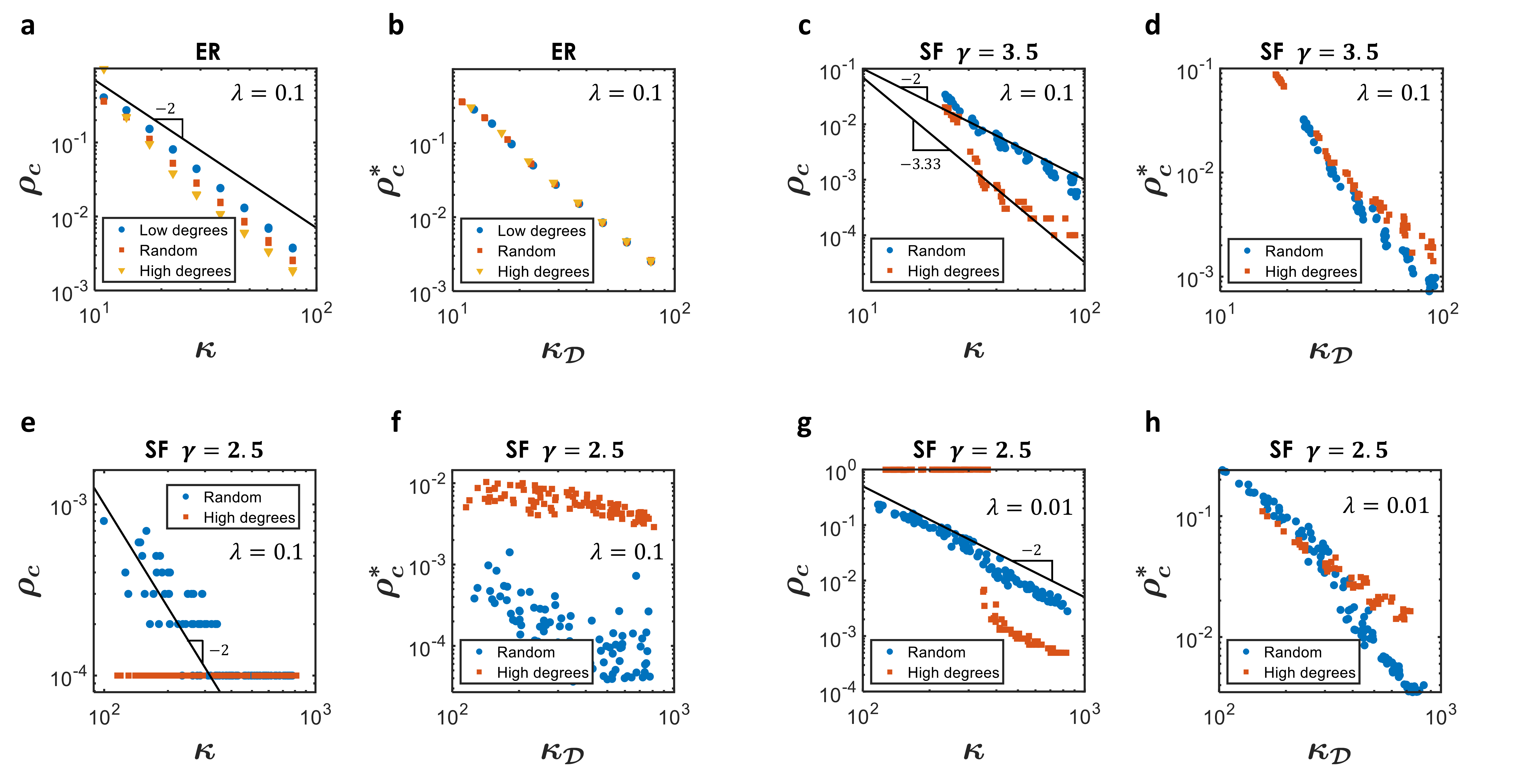}
	\caption{ 
		\textbf{Selection methods of the controlled nodes.} 
		(a) For cellular dynamics on ER network, we show the results for three different ways to select the controlled nodes, highest degree nodes, lowest degree nodes and randomly. The scaling of the minimal fraction for sustaining, $\rho_c$, as $\kappa$ is similar, but the value of $\rho_c$ changes remarkably. The interaction strength is $\lambda=0.1$. $\rho_c$ was calculated for 10 ER networks of $N=10^4$ for each mean degree $\av{k}$. The parameters of the cellular dynamics were set to $a=1,\ h=2$.
		(b) According to Eq.\ \eqref{ForcedMFArbitraryBeta}, for arbitrary degree distribution of the controlled nodes, the relevant parameters are $\beta_{\D}$ and $\rho^*$ instead of $\beta$ and $\rho$. Indeed, in these variables the three cases yield a single curve. The data points are the same as in (a), shown with respect to the new variables $\beta_{\D}$ and $\rho^*$. 				
		(c)-(d) Results for scale-free networks with $\gamma=3.5$ and varying minimal degree $k_0$. Here the scaling of $\rho_c$ with $\beta$ changes between high degrees and random selections, as predicted by Eq.\ \eqref{ScalingRhocBetaSFHighMM} for high degrees selection and Eq.\ \eqref{RhocMMScaling} for a random selection. (d) The two selections are close to be on a single curve, but it is much less significant compared to ER in (b) since the results for SF are less accurate because of the larger fluctuations.
		(e)-(f) Results for SF with exponent $\gamma=2.5$. This network is more challenging for the mean-field theory since the variation of node degrees is larger. The exponent -2 for random selection still shows a good agreement with simulations. For high degree selection we see a single node sustaining, since the network size is $10^4$. 
		(f) Our theory does not cover this challenging case, particularly because we reach the limit of microscopic sustaining. Thus, we decreased $\lambda$ to be $0.01$.
		(g)-(h) The same as (e)-(g) with smaller interaction strength $\lambda=0.01$. Here we do not see anymore a single node sustaining. The exponent -2 for random selection still works. However, our theory does not capture the behavior for high degrees selection. 
	}
	\label{fig:Selection}	
\end{figure}

In cellular dynamics as shown in Eq. \eqref{RhocMMScaling} there exists a power-law relation $\rho_c\sim\beta^{-\theta}$ where $\theta=h/(h-a)$. Thus, for high degrees selection of the forced nodes, we use Eq.\ \eqref{ScalingRhoBetaArb} to get
\begin{equation} \label{ScalingRhocBetaSFHighMM}
	\rho_c \sim \beta^{- \frac{h}{h-a} \frac{\gamma-1}{\gamma-2}}.
\end{equation}
For $a=1,\ h=2$ and $\gamma=3.5$, we get the scaling $\rho_c\sim\beta^{\frac{10}{3}}$ for controlled nodes with high degrees, and $\rho_c\sim\beta^{-2}$ for random selection as shown in Fig.\ \ref{fig:Selection}c with qualitatively agreement. We do not expect a perfect agreement since our MF of small fluctuations has deviations from the simulations in SF networks as shown in Fig.\ 4a,b in the main text.

\clearpage

\subsection{Neuronal dynamics}

We consider a modified Wilson-Cowan model \cite{wilson1972excitatory,wilson1973mathematical} for excitation in neuronal networks,
\begin{equation}
\dod{x_i}{t} = -x_i + \lambda \sum_{j = 1}^N \m Aij \dfrac{1}{1 + e^{\mu-\delta x_j}},
\label{Neuronal}
\end{equation}
which we examine for $\mu = 5$ and $\delta = 1$.
In this version of the model, adapted to the form of Eq.\ (\ref{Dynamics}), the summation is extracted outside of the exponential function, namely we write $\lambda \sum_{j = 1}^N \m Aij (1 + e^{\mu-\delta x_j})^{-1}$ instead of $(1 + e^{\mu-\delta \lambda \sum_{j = 1}^N \m Aij x_j})^{-1}$. As a result, each node receives accumulating inputs from all its neighbors, and hence higher in-degree nodes gain a stronger overall activation signal from their surrounding neighbors. This is in contrast to the standard version of the model, with the summation appears inside the exponent, and hence the effect of the multiple incoming signals quickly reaches saturation. In the present context, where we wish to observe the role of degree heterogeneity on reigniting, \textit{e.g}., through $\kappa$, the form of Eq.\ (\ref{Neuronal}) provides a more relevant testing ground.


\subsubsection{Free system}	

To obtain the fixed-points of the system we use Eq.\ \eqref{BetaXFree} to obtain the relation,
\begin{equation} \label{betaXBrain}
\beta = \bar{\x} (1+e^{\mu -\delta \bar{\x}}).
\end{equation}
Plotting $\bar{\x}$ vs.\ $\beta$ (Fig.\ 5b in the main text) we obtain the three dynamic phases of the free system. The \textit{inactive} state $\x_0$ (red), in which all activities are suppressed, is obtained when the network is extremely sparse, \textit{i.e}.\ small $\beta$. The \textit{active} $\x_1$ (green), in which $x_i$ are relatively high, is observed when $\beta$ is large. In between these two extremes, the system features a \textit{bi-stable} phase, in which both $\x_0$ and $\x_1$ are potentially stable. These phases are separated by two critical points $\beta_{c,1} < \beta_{c,2}$, between which the system is bi-stable and thus unsustainable. However, above $\beta_{c,2}$ the system is naturally sustainable without any external intervention.
To find $\beta_{c,12}$ we look for the extremum points of $\beta$ vs.\ $\bar{\x}$ in Eq.\ \eqref{betaXBrain}, getting
\begin{equation} 
	 1+e^{\mu -\delta \bar{\x}} + \bar{\x} (-\delta e^{\mu -\delta \bar{\x}}) = 0,
\end{equation}
that turns to
\begin{equation} 
	e^{\delta \bar{\x}-\mu}+1 = \delta\bar{\x} .
\end{equation}
This equation is transcendental but solvable numerically with two solutions $\bar{\x}_{c,1}$ and $\bar{\x}_{c,2}$, providing through Eq.\ \eqref{betaXBrain} the borders of the three phases of a free neuronal system, inactive ($\beta<\beta_{c,1}$), unsustainable ($\beta_{c,1} \le \beta \le \beta_{c,2}$) and naturally sustainable ($\beta>\beta_{c,2}$).

\subsubsection{Forced system}

To obtain the states of the system when a random fraction of nodes is controlled with high activity $\Delta$, we use Eqs.\ \eqref{BetaXForced} and \eqref{Neuronal} to get
\begin{equation} \label{betaXBrainForced}
	\beta = \dfrac{\bar{\x}}{\dfrac{1-\rho}{1+e^{\mu -\delta \bar{\x}}}+\dfrac{\rho}{1+e^{\mu -\delta \Delta}}} .
\end{equation}
In Fig.\ 5b in the main text we present this result which agrees well with simulations results, and shows a new phase diagram for the forced system. Even though both forced and free system for brain dynamics show an s-shape diagram, the diagram of the forced system is biased to the left, and thus has a significantly smaller $\beta_{c,2}$. This creates a window (blue shade) of sustainability where the control drives the system to become sustainable.

\subsubsection{Sustaining a network}

For finding $\beta_c$ above which a forced system becomes sustainable, we again find the local maximum of $\beta$ vs.\ $\bar{\x}$ in Eq.\ \eqref{betaXBrainForced}, yielding
\begin{equation} 
	\dfrac{1-\rho}{1+e^{\mu -\delta \bar{\x}}}+\dfrac{\rho}{1+e^{\mu -\delta \Delta}} - \bar{\x} \dfrac{1-\rho}{\left(1+e^{\mu -\delta \bar{\x}}\right)^2} \delta e^{\mu -\delta \bar{\x}} = 0.
\end{equation}
Solving numerically this transcendental equation, we get $\bar{\x}_{c,1}$ and $\bar{\x}_{c,2}$, and using Eq.\ \eqref{betaXBrainForced} we obtain $\beta_{c,1}$ and $\beta_{c,2}$ for the edges of the unsustainable range of a forced system. The higher value among them is the critical $\beta_c(\rho,\Delta)$ required for sustaining the network. This result is presented in Fig.\ 5c,d in the main text, showing a good agreement with simulations.

\subsubsection{Tricritical point}

For finding $\rho_0$, above which there is no unsustainable region, we have to apply the condition of Eq.\ \eqref{Maximum2} on Eq.\ \eqref{betaXBrainForced}. Together with the condition of Eq.\ \eqref{Maximum} on Eq.\ \eqref{betaXBrainForced}, and Eq.\ \eqref{betaXBrainForced} itself, we obtain numerically the tricritical point $(\beta_0,\rho_0)$, beyond which there is no unsustainable region as can be seen in Fig.\ 5d in the main text.

\subsection{Spin dynamics}

Here we explore the last example, spin dynamics. We consider a kinetic model based on Ising-Glauber model \cite{krapivsky2010kinetic} to follow the state of spins connected by a network of ferromagnetic interactions.

\subsubsection{Free system}

The dynamics of the spins according to this model \cite{krapivsky2010kinetic} is captured by the set of equations which are not included in Eq.\ \eqref{Dynamics},
\begin{equation}
	\dod{x_i}{t} = 
	-x_i +  \tanh \bigg( \lambda \sum_{j = 1}^N \m Aij x_j \bigg).
	\label{Spins}
\end{equation}
Even though this form is beyond our general framework in \eqref{Dynamics}, still a similar analysis can be performed as follows.
First, we look on a relaxed system, \textit{i.e.}\ $\dif x_i/ \dif t=0$, giving
\begin{equation}	
x_i^* = \tanh \bigg( \lambda \sum_{j = 1}^N \m Aij x_j^* \bigg).
\end{equation}
The next step is the DBMF approximation \cite{PastorSatorras2001prl,boguna2002epidemic,Barrat2008,Dorogovtsev2008,PastorSatorras2015} that we used above, Eq.\ \eqref{DBMF1}, according which we substitute $\m Aij$ by $k_ik_j/(N\av{k})$,
\begin{equation}	
	x_i^* = \tanh \bigg( \lambda k_i \sum_{j = 1}^N \frac{k_j}{N\kk} x_j^* \bigg).
\end{equation}
As above, Eq.\ \eqref{XFree}, we use the average $\bar{\x}$ to write
\begin{equation}	
	x_i^* = \tanh \bigg( \lambda k_i \bar{\x} \bigg).
\end{equation}
Next, we insert the last equation into the definition of $\bar{\x}$ to obtain a self-consistent equation for $\bar{\x}$,
\begin{equation}	
	 \bar{\x} = \sum_{i = 1}^N \frac{k_i}{N\kk} \tanh \bigg( \lambda k_i \bar{\x} \bigg).
\end{equation}
Replacing the summation on nodes $i$ by summation on the degrees $k$, we obtain
\begin{equation}	
	\bar{\x} = \sum_{k } \frac{kp_k}{\kk} \tanh \bigg( \lambda k \bar{\x} \bigg)
\end{equation}	
which is a single solvable theoretically equation given the degree distribution.
As above, Eq.\ \eqref{InsAve1}, as in Ref.\ \cite{Gao2016} and explained above in Section \ref{SecMbarxRbarx}, we insert the average into the function for small fluctuations and/or if $\lambda k \bar{\x}$ values are close to zero were $\tanh$ is close to linear, or $\lambda k \bar{\x}$ values are large where $\tanh$ is close to constant. Using this approximation, we obtain 
\begin{equation}	
	\bar{\x} = \tanh ( \beta \bar{\x} ),
	\label{XandBetaSpins}
\end{equation}
which predicts the states of a free system of spin dynamics. The phase diagram constructed from this equation is shown in Fig.\ 5f (thin light lines) in the main text.
Isolating $\beta$ in Eq.\ \eqref{XandBetaSpins} yields a simpler equation,
\begin{equation}
	\beta = \frac{ \tanh^{-1} (\bar{\x})}{ \bar{\x} } = \frac{1}{2\bar{\x}}  \ln \bigg( \frac{1+\bar{\x}}{1-\bar{\x}} \bigg) .
	\label{BetaXSpins}
\end{equation}

\subsubsection{Forced system}

Here we follow the steps we have done above in Section \ref{SecForced}. When controlling the set $\F$ to have an activity $\Delta$, we have to separate the sum in Eq.\ \eqref{Spins} between neighbors in $\D$ in $\F$. Thus, the equation for node $i\in\D$ becomes
\begin{equation}
	\dod{x_i}{t} = 
	-x_i +  \tanh \bigg( \lambda \sum_{j \in \D} \m Aij x_j + \lambda \sum_{j \in \F} \m Aij \Delta \bigg).
\end{equation}
In relaxation,
\begin{equation}
	x_i^* = \tanh \bigg( \lambda \sum_{j \in \D} \m Aij x_j^* + \lambda k_i^{\D\to\F} \Delta \bigg).
\end{equation}
where $k_i^{\D\to\F}=\sum_{j\in\F}\m Aij$ is the number of neighbors in $\F$ of the node $i$ belonging to $\D$. Next, we perform again the DBMF approximation \cite{PastorSatorras2001prl,boguna2002epidemic,Barrat2008,Dorogovtsev2008,PastorSatorras2015}, replacing $\m Aij$ by the probability of being a link $(i,j)$ given their relevant degrees, which here are degrees inside $\D$. Hence, 
\begin{equation}
	x_i^* = \tanh \bigg( \lambda k_i^{\D\to\D} \sum_{j \in \D} \frac{k_j^{\D\to\D}}{|\D|\av{k_{\D\to\D}}} x_j^* + \lambda k_i^{\D\to\F} \Delta \bigg).
\end{equation}
Using the same definition as above of $\bar{\x}$, Eq.\ \eqref{XForced}, we obtain
\begin{equation}
	x_i^* = \tanh \Big( \lambda k_i^{\D\to\D} \bar{\x} + \lambda k_i^{\D\to\F} \Delta \Big).
\end{equation}
Plugging the last equation into the definition of $\bar{\x}$, we get a self-consistent equation for $\bar{\x}$,
\begin{equation}
	\bar{\x} = \sum_{j \in \D} \frac{k_j^{\D\to\D}}{|\D|\av{k_{\D\to\D}}} \tanh \Big( \lambda k_j^{\D\to\D} \bar{\x} + \lambda k_j^{\D\to\F} \Delta \Big) .
\end{equation}
Next, we use again the mean-field approximation \cite{Gao2016} of moving the averaging into the function $\tanh$, as above Eq.\ \eqref{ThetaXForced}, Section \ref{SecMbarxRbarx} and Eq.\ \eqref{XandBetaSpins}, to obtain 
\begin{equation}
	\bar{\x} = \tanh \Big( \lambda \kappa_{\D\to\D} \bar{\x} + \lambda \kappa_{\D\to\F} \Delta \Big) ,
\end{equation}
where $\kappa_{\D\to\D}$ and $\kappa_{\D\to\F}$ are defined in Eq.\ \eqref{kappaDDkappaDF} and found for a random selection in Eqs.\ \eqref{kappaDD} and \eqref{kappaDF}. Substituting their values for a random selection of the forced nodes, yields
\begin{equation}
	\bar{\x} = \tanh \Big( \lambda (1+(\kappa-1)(1-\rho)) \bar{\x} + \lambda (\kappa-1)\rho \Delta \Big) ,
\end{equation}
which for large $\kappa$ becomes
\begin{equation}
	\bar{\x} = \tanh \big( \beta (1-\rho) \bar{\x} + \beta\rho \Delta \big) .
	\label{XandBetaForcedSpins}
\end{equation}
This equation predicts the states of the forced system as shown in Fig.\ 5f in the main text. A simpler relation can be obtained by isolating $\beta$,
\begin{equation}
	\beta = \frac{ \tanh^{-1} (\bar{\x})}{(1-\rho) \bar{\x} + \rho \Delta } = \frac{1}{2}  \frac{  \ln((1+\bar{\x})/(1-\bar{\x}))}{(1-\rho) \bar{\x} + \rho \Delta }  .
	\label{BetaXForcedSpins}
\end{equation}

\clearpage

\bibliographystyle{unsrt}
\bibliography{bibliography}